\documentclass[conference]{IEEEtran}

\usepackage[square,numbers,sort]{natbib}
\usepackage{xcolor}
\usepackage{amsmath}
\usepackage{amssymb}
\usepackage{hyperref}
\usepackage{tikz}
\usepackage{xspace}
\usepackage{subcaption}
\usepackage{tabularx}
\usepackage{longtable}
\usepackage{booktabs}
\usepackage{multirow}
\usepackage{colortbl}
\usepackage{pifont}
\usepackage[most]{tcolorbox}
\usepackage{float}
\usepackage{enumitem}
\usepackage{xfp}  
\usepackage{wheelchart}
\usepackage{etoolbox}
\usepackage{pgfplots}
\usepackage{pgfplotstable}
\usepackage{arydshln}  
\usepackage{array}
\usepackage{listings}
\usepackage{balance}
\usetikzlibrary{patterns,patterns.meta}
\newcolumntype{L}[1]{>{\raggedright\arraybackslash}p{#1}}
\newcolumntype{C}[1]{>{\centering\arraybackslash}p{#1}}

\definecolor{bannerColor}{HTML}{ab4b52}
\definecolor{noBannerColor}{HTML}{0072bb}

\definecolor{firstparty}{HTML}{78c895}
\definecolor{thirdparty}{HTML}{ba9287}
\definecolor{ads}{HTML}{5abfe7}

\definecolor{beforeclick}{HTML}{1794E8}
\definecolor{afterclick}{HTML}{E86B17}

\definecolor{Blue}{HTML}{4169e1}
\definecolor{Brown}{HTML}{cf6161}
\definecolor{Green}{HTML}{008000}
\definecolor{Violet}{HTML}{ee82ee}
\definecolor{Red}{HTML}{ff0000}
\definecolor{Plum}{rgb}{0.58, 0, 0.647}
\definecolor{Teal}{HTML}{008080}

\definecolor{Magenta}{HTML}{ff00ff}
\definecolor{Crimson}{HTML}{700c1a}

\definecolor{lightgreen}{rgb}{0.8, 1, 0.8}
\definecolor{midgreen}{rgb}{0.4, 0.8, 0.4}
\definecolor{darkgreen}{rgb}{0.0, 0.5, 0.0}

\definecolor{lightblue}{HTML}{6abce2}
\definecolor{lightred}{HTML}{ffbead}

\newcommand{\heatmap}[1]{%
	\ifdim #1 pt < 0pt
	\cellcolor{white!\fpeval{(100 + #1)}!lightblue}{#1}%
	\else
	\cellcolor{white!\fpeval{(100 - #1)}!lightred!70}{#1}%
	\fi
}

\newcommand{\heatmaps}[1]{%
	\ifdim #1 pt < 0pt
	\cellcolor{white!\fpeval{100*(1 + #1)}!lightblue}{#1}%
	\else
	\cellcolor{white!\fpeval{100*(1 - #1)}!lightred!70}{#1}%
	\fi
}

\newcommand{\heatmappeachy}[1]{%
	\cellcolor{white!\fpeval{100*(1 - #1^3)}!peach!70}{#1}%
}

\definecolor{fig-yellow}{HTML}{f1c233}
\definecolor{fig-pink}{HTML}{eeeeee}
\providecommand{\sysname}{RegTrack\xspace}
\providecommand{\AT}{\emph{A\&T}\xspace}

\newcommand{\eg}{e.g.,\@\xspace}

\newcommand{\etc}{etc.\@\xspace}

\newcommand{\noindgras}[1]{\noindent{\bf #1}}

\newcommand{\sectionref}[1]{$\S$\ref{#1}}

\def\Snospace~{\S{}}

\definecolor{peach}{RGB}{255, 218, 185}

\begin{document}
\date{}

\title{\sysname: Uncovering	Global Disparities in Third-party Advertising and Tracking}

\author{
  \IEEEauthorblockN{Tanya Prasad \qquad Rut Vora \qquad Soo Yee Lim \qquad Nguyen Phong Hoang \qquad Thomas Pasquier}
  \IEEEauthorblockA{The University of British Columbia, Vancouver, Canada}
}

\IEEEoverridecommandlockouts
\makeatletter\def\@IEEEpubidpullup{3\baselineskip}\makeatother
\IEEEpubid{%
\parbox{\columnwidth}{
    {\fontsize{7.5}{7.5}\selectfont Workshop on Measurements, Attacks, and Defenses for the Web (MADWeb) 2026 \\
    27 February 2026, San Diego, CA, USA \\
    ISBN 978-1-970672-06-0 \\
    https://dx.doi.org/10.14722/madweb.2026.23010 \\
    www.ndss-symposium.org}
}
\hspace{\columnsep}\makebox[\columnwidth]{}}

\maketitle

\begin{abstract}
Third-party advertising and tracking (\AT) are pervasive across the web, yet user exposure varies significantly with browser choice, browsing location, and hosting jurisdiction.
We systematically study how these three factors shape tracking by conducting synchronized crawls of 743 popular websites from 8 geographic vantage points using 4 browsers and 2 consent states.
Our analysis reveals that browser choice, user location, and hosting jurisdiction each shape tracking exposure in distinct ways.
Privacy-focused browsers block more third-party trackers, reducing observed \AT domains by up to 30\% in permissive regulatory environments, but offer smaller relative gains in stricter regions.
User location influences the tracking volume, the prevalence of consent banners, and the extent of cross-border tracking: GDPR-regulated locations exhibit about 80\% fewer third-party \AT domains before consent and keep 89--91\% of \AT requests within the EEA or adequacy countries.
Hosting jurisdiction plays a smaller role; tracking exposure varies most strongly with inferred user location rather than where sites are hosted.
These findings underscore both the power and limitations of user agency, informing the design of privacy tools, regulatory enforcement strategies, and future measurement methodologies.
\vspace{-.44cm}
\end{abstract}

\IEEEpeerreviewmaketitle

\section{Introduction}
\label{sec:introduction}
Third-party advertising and tracking (\AT) underpin much of the web's business model, with these technologies present across the web ecosystem~\cite{urban2020beyond,bujlow2017survey,Demir2022TowardsUF}.
These mechanisms enable cross-site profiling, where advertisers and intermediaries infer browsing behavior, interests, and demographics from identifiers and interaction traces~\cite{mayer2012third,bujlow2017survey}.
Browsers exercise substantial control over user privacy by mediating page execution and network requests, exposing levers to shape storage and communication~\cite{roesner2012sharemenot,mayer2012third}.
This technical position allows browsers to block or rewrite requests, restrict cookies and other state, and deploy anti-fingerprinting and anti-tracking defenses~\cite{iqbal2022khaleesi,siby2021zubair}.
However, dominant browsers are developed by firms with significant advertising businesses, raising questions about incentive alignment between privacy protections and advertising addressability, motivating empirical evaluation of what users actually experience.

Different jurisdictions have introduced privacy regulations constraining tracking and data flows, with varying obligations and enforcement models~\cite{degeling2019gdpr,matte2020tcf,hosseini2024ccpa}.
The EU's GDPR emphasizes opt-in consent for tracking, whereas US laws like CCPA and CPRA center opt-out from ``sale'' or ``sharing''~\cite{degeling2019gdpr,hosseini2024ccpa}.
Empirical audits report substantial variation in compliance and dark patterns in consent interfaces, producing geographic disparities in effective privacy protection~\cite{matte2020tcf,santos2020cmp,Tran2024MeasuringCW,Tran2025DarkPI}.

These observations point to a methodological gap: prior work has not systematically compared the \emph{relative} and \emph{joint} effects of \emph{browser choice}, \emph{user location}, and \emph{hosting jurisdiction} on tracking exposure using a single, controlled measurement design.
Cookie banners and vendor responses may differ for EU vs.\ non-EU users~\cite{Eijk2019LocationCookieNotices,Degeling2019GDPRImpact}; trackers may react to IP-level geo signals~\cite{Mishra2020DontCountMeOut,Tran2024MeasuringCW}; and sites may deploy jurisdiction-wide policies based on where they host.
We therefore ask three interconnected questions, ordered by decreasing user agency:

\begin{itemize}[leftmargin=*]
    \item \textbf{RQ1: Effect of browsers}---How do different browsers affect third-party \AT\ exposure?
    \item \textbf{RQ2: Effect of user location}---How does a user's geographic location (as inferred by sites) influence \AT?
    \item \textbf{RQ3: Effect of hosting location}---How does a website's hosting jurisdiction affect third-party \AT\ practices?
\end{itemize}

We begin with browser choice (RQ1), which users control directly; proceed to user location (RQ2), which users can sometimes influence (e.g., via VPNs); and conclude with hosting jurisdiction (RQ3), a structural factor beyond individual control.
This ordering lets us assess how each lever contributes to observed tracking exposure under matched conditions.
To support this, we design \sysname, a consent-aware factorial measurement framework varying browser and browsing location over a shared list of popular sites, using synchronized crawls from 8 vantage points and attributing third-party requests to known \AT\ domains.
Our main contributions are:
\begin{itemize}[leftmargin=*]
    \item We design \sysname, a multifactor, consent-aware measurement framework systematically studying the effects of browser, user location, and website hosting jurisdiction on third-party \AT.
    \item We collect and analyze a dataset spanning 8 geographic vantage points, 4 browsers, and 743 popular websites, enabling within-factor contrasts and interaction analysis.
    \item We compare the magnitude of browser-, location-, and hosting-related differences in tracking exposure and identify where user-controllable choices provide meaningful leverage versus where structural forces dominate.
\end{itemize}

Our measurements yield three main findings.
First, browser choice matters most in permissive environments: privacy-focused browsers (e.g., Brave) substantially reduce tracking in the US and opt-out contexts, while differences narrow in stricter regions like the EU.
Second, user location has a large effect on baseline tracking and additional tracking unlocked after clicking ``Accept,'' with EU vantages showing lower pre-consent exposure but large post-consent jumps.
Third, hosting jurisdiction plays a secondary role: most observed discrimination in banners and tracking behavior is driven by inferred user location rather than website hosting location.

\begin{figure*}[htpb]
	\centering
	\includegraphics[width=0.9\textwidth]{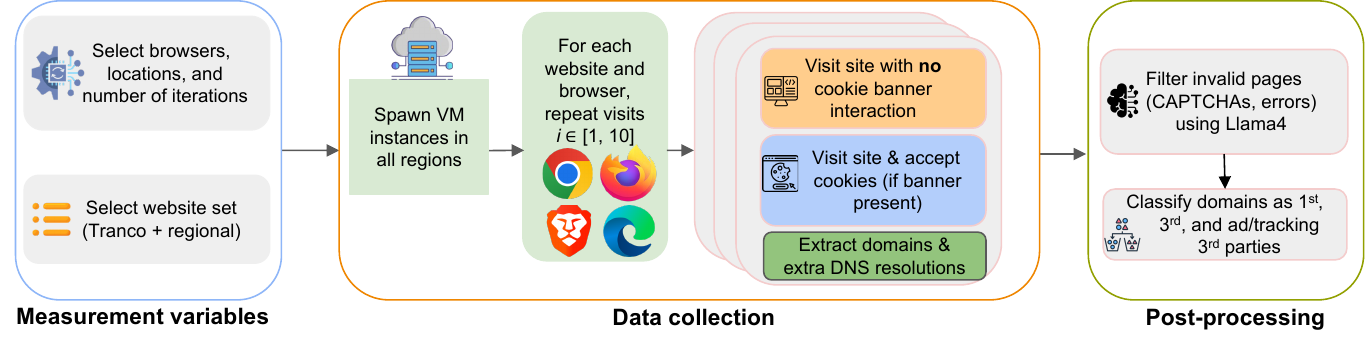}
	\caption{
	Overview of \sysname.
	We fix browsers, locations, and consent states, then crawl websites from per-location VMs, spawning containers for each (browser, site, consent) configuration to collect HAR files and screenshots.
	Post-processing filters invalid pages with a multimodal LLM and labels domains as 1st-party, other 3rd-party, or \AT 3rd-party using public blocklists.
	}
	\label{fig:framework}
\vspace{-.5cm}
\end{figure*}

\section{Background and Motivation}
\label{sec:motivation}
The web economy relies on embedded 3rd-party services for advertising, analytics, and personalization, enabling cross-site identification and profiling at scale~\cite{mayer2012third,englehardt2016online,bujlow2017survey}.
These services increasingly use request- and redirect-based techniques and 1st-party integrations rather than only classic 3rd-party cookies~\cite{urban2020beyond,iqbal2022khaleesi}.
As our goal is to \emph{attribute} differences across browser choice, user location, and hosting jurisdiction, we focus on network-visible outcome measures comparable across factors.

We define \emph{tracking exposure} as the set of 3rd-party apex domains contacted during page loads, labeled as \AT or ``other'' using curated public lists, along with the prevalence of consent banners and changes between pre- and post-consent conditions~\cite{urban2020beyond}.
This definition favors coverage and comparability while acknowledging that some behaviors (payload content, fingerprinting) are not directly observable at network level.
Since user studies show 72\% choose ``Accept all''~\cite{habib2022okay}, we treat \emph{no-click} and \emph{accept-all} as separate experimental states.

\emph{Browser choice} is the lever users control most directly; mainstream browsers differ in default protections (3rd-party cookie restrictions, storage partitioning, anti-fingerprinting) that reshape request-level exposure~\cite{madhusudhan2022privacy}.
\emph{User location} conditions banner presentation and when 3rd-party requests are initiated, with field studies documenting location-dependent consent surfaces~\cite{degeling2019gdpr,hosseini2024ccpa}.
\emph{Hosting jurisdiction} shapes data recipients and applicable legal regimes; large-scale mapping shows cross-border transfers are common~\cite{iordanou2018tracing,urban2020beyond}.
\sysname manipulates these three factors within a single framework, enabling contrasts respecting their user agency ordering.

Prior work typically vary one factor at a time or rely on setups hard to compare across papers~\cite{cassel2022omnicrawl,degeling2019gdpr,iordanou2018tracing}.
Our 4x8 factorial design with two consent states reveals how \emph{browser choice}, \emph{user location}, and \emph{hosting jurisdiction} each affect 3rd-party \AT exposure when other factors are held constant.

\section{Measurement Methodology}
\label{sec:methodology}
To compare how browser, user location, and hosting jurisdiction shape third-party \AT, we built \sysname, a consent-aware factorial framework concurrently crawling websites from eight vantage points using four browsers and two consent states (\emph{no-click} and \emph{accept-all}) (\autoref{fig:framework}).
Each (browser, location, site, consent) configuration runs in a fresh container, recording HTTP requests to HAR files and screenshots.

\subsection{Measurement Variables}
\label{subsec:measurement-variables}

\noindgras{Browsers.}
We evaluate four mainstream desktop browsers with distinct privacy postures: Chrome, Edge, Firefox, and Brave, covering market-dominant and privacy-focused options~\cite{statcounterDesktopBrowser,madhusudhan2022privacy} (market shares: 65.54\%, 13.89\%, 6.36\%, and 1\%).
All crawls run on Linux in default configuration without extensions to capture ``out of the box'' exposure~\cite{Habib2021EvaluatingTU}.

\noindgras{Browsing locations.}
We deploy crawls from eight regions: California (USA), Ohio (USA), Quebec (Canada), Mumbai (India), Singapore, Frankfurt (Germany), Paris (France), and Dublin (Ireland), covering major privacy regimes including GDPR~\cite{GDPR2016}, CCPA~\cite{CCPA2018}, PDPA~\cite{PDPA2012}, PIPEDA~\cite{PIPEDA2000}, and DPDPA~\cite{DPDPA2023}.
We treat these locations as representative bundles of legal obligations, enforcement practices, and consent norms, interpreting results at broad regulatory cluster levels (e.g., GDPR vs.\ opt-out frameworks).

\noindgras{Websites.}
To avoid regional bias, we combine globally and regionally popular sites: top-1K Tranco~\cite{letranco} augmented with top-100 per country (US, IN, SG, DE, FR, IE) from Cloudflare Radar~\cite{cloudflare_radar}, filtered to top-100K Tranco ranking, yielding 1{,}005 unique domains.
Five additional domains (\texttt{ovh.net}, \texttt{hotstar.com}, \texttt{truecaller.com}, \texttt{swiggy.com}, \texttt{google.ie}) are from the country lists.

\subsection{Crawling Architecture}
\label{subsec:crawling-setup}

Our crawler is built on Browsertime~\cite{browsertime}, running each visit in a fresh Docker container to ensure no client-side state persists.
For each configuration, \sysname loads the page, records network requests (HAR), and captures a screenshot.
We focus analysis on apex domains (stable across visits) while using FQDNs for blocklist matching.\footnote{E.g., we classify \texttt{ads.google.com} rather than \texttt{google.com}.}
We visit each site 10 times per configuration; apex counts converge after roughly five visits, with additional visits providing robustness against transient failures.

\subsection{Cookie Consent Handling}
\label{subsec:cookie-consent-handling}

Cookie-consent banners can gate content access and substantially change which third-party requests are issued~\cite{degeling2019gdpr}.
For each site and configuration, \sysname performs two independent passes (repeated 10 times): a \emph{no-click} pass without consent interface interaction, and an \emph{accept-all} pass selecting the most affirmative option when a banner is present.
We choose the most affirmative option as it provides an upper bound on tracking exposure and represents realistic user behavior (72\% select ``OK'' or equivalent~\cite{habib2022okay}).

\noindgras{Banner interaction.}
We automate ``accept'' clicks using CMP-specific selectors (Didomi, Quantcast, OneTrust, CookieBot) with text-based heuristic fallback for affirmative labels from a curated lexicon.\footnote{We manually curated the lexicon by sampling sites with banners, collecting acceptance strings, and translating them into represented languages.}
The detection logic is injected via Browsertime's JavaScript hooks and executed across all iframes.
We manually audited a sample of pages to remove lexicon entries producing false positives.
\sysname successfully interacts with banners on 91--95\% of banner-using sites; remaining cases are treated as no-click.

\subsection{Data Cleaning}
\label{subsec:data-cleaning}

We exclude visits where the intended page did not load due to CAPTCHAs, block pages, or network errors.
To detect such cases at scale, we classify screenshots using open-source vision-language models.
After evaluating several models against 400 manually labeled screenshots (\autoref{appendix:section-c}), we select Llama4, which achieves 99\% accuracy (Cohen's $\kappa\!=\!0.94$).
A site is included if it loads successfully in $\geq5/10$ visits for every (browser, location, consent) configuration.
Of the initial 1{,}005 domains, 26\% fail this criterion for at least one configuration (often due to CAPTCHAs or geo-blocking).
Our final cleaned dataset contains 743 sites.

\subsection{Domain Classification}
\label{subsec:domain-classification}

From each HAR file, we extract all requested domains and classify them.
A domain is \emph{first-party} if its apex matches the visited site's pay-level domain; otherwise, it is \emph{third-party}.
Third-party domains are classified as \AT or ``other'' using a union of widely used public blocklists (EasyList/EasyPrivacy~\cite{easylist-default,Easylist-privacy},
AdGuard~\cite{adguard-default}, and others; full list in \autoref{appendix:blocklists}).
If a domain appears on any list, we flag it as potentially \AT-related; remaining third-party domains are labelled ``other.''
Since these blocklists are primarily designed for URL-level filtering, our domain-level matching may introduce measurement error; we discuss this limitation and our mitigation in~\sectionref{disc:limitation}.

\section{Data Analysis and Results}
\label{sec:analysis}
We now present our empirical results, organized around the three levers of user agency and the research questions in~\autoref{sec:introduction}.

\subsection{Browser Choice}
\label{subsec:results-browser}

\begin{figure}[t]
\centering
\begin{tikzpicture}
\begin{axis}[
width=\columnwidth,
height=5cm,  
ybar stacked,
bar width=15pt,
x=2cm,  
enlarge x limits=0.15,
xtick      = {0,1,2,3},
xticklabels={Brave,Chrome,Edge,Firefox},
xtick=data,
axis x line*=bottom,
axis y line*=left,
ymin=0,  
scaled y ticks=false,
ytick = {0,10000,20000,30000},
yticklabels={0,10k,20k,30k},
ylabel={}, 
xlabel={},
legend style={
    at={(0.5,1.05)},
    anchor=north,
    legend columns=3,  
    draw=none,
    fill=none,
    /tikz/every even column/.append style={column sep=0.5cm},
},
legend cell align={left},
area legend,  
]

\makeatletter
\newcommand\resetstackedplots{%
\pgfplots@stacked@isfirstplottrue
\addplot [forget plot,draw=none] coordinates {(0,0) (1,0) (2,0) (3,0)};
}
\makeatother

\addplot+[bar shift=-8pt, fill=firstparty, draw opacity=0, area legend] coordinates {(0,742) (1,742) (2,742) (3,738)};
\addplot+[bar shift=-8pt, fill=thirdparty, draw opacity=0, area legend] coordinates {(0,5614) (1,7340) (2,7342) (3,6761)};
\addplot+[bar shift=-8pt, fill=ads, draw opacity=0, area legend] coordinates {(0,15921) (1,23219) (2,22759) (3,20997)};

\resetstackedplots

\addplot+[bar shift=+8pt, draw=none, forget plot,
pattern={Lines[angle=45,distance=10pt,line width=8pt]},
pattern color=firstparty] coordinates {(0,742) (1,742) (2,742) (3,736)};
\addplot+[bar shift=+8pt, draw=none, forget plot,
pattern={Lines[angle=45,distance=10pt,line width=8pt]},
pattern color=thirdparty] coordinates {(0,4651) (1,5008) (2,4969) (3,4877)};
\addplot+[bar shift=+8pt, draw=none, forget plot,
pattern={Lines[angle=45,distance=10pt,line width=8pt]},
pattern color=ads] coordinates {(0,12528) (1,14374) (2,14067) (3,12841)};

\legend{First party, Third party, \emph{A\&T}}

\end{axis}
\end{tikzpicture}
\caption{Total number of \AT, by browser, after accepting all cookies in US-Ohio (plain) and France (striped).}
\label{fig:post_cookie_ohio_france}
\vspace{-0.5cm}
\end{figure}
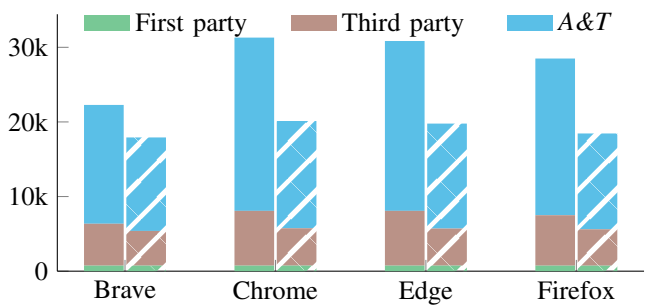

Browser choice represents the factor over which users exercise the most direct control.
Unlike user location or website hosting jurisdiction, users can freely select and switch between browsers with minimal technical barriers, legal restrictions, or geographic constraints.
However, users typically commit to a single browser due to ecosystem lock-in and familiarity, making this choice particularly consequential for their long-term tracking exposure \cite{mozillaBeyondChoice}.

To understand how browser choice affects the number of third-party \AT domains, we compare two regulatory contexts: USA--Ohio, with relatively permissive privacy regulations, and France, governed by the GDPR.
In USA--Ohio, Brave offers the lowest exposure, triggering 31\% fewer \AT domains\footnote{Our \AT classification relies on domain-level blocklist matching, which may over- or under-count tracking in some cases; see \sectionref{disc:limitation} for details.} than Chrome, the browser with the highest tracking levels, as illustrated in~\autoref{fig:post_cookie_ohio_france}.
On the other hand, the differences between browsers are less pronounced in France.
While Brave still yields the lowest exposure, triggering 13\% fewer \AT domains than Chrome, the gap between browsers narrows considerably in this GDPR-regulated region.

Browser choice significantly affects third-party \AT exposure, but other factors, specifically user location itself, appear to play an important role.
We therefore turn to our second research question, examining how user location shapes tracking practices independent of browser selection.

\subsection{Effect of User Location}
\label{subsec:results-location}

We now turn to user location, a factor users can only partially influence (for example, via VPNs) but which determines the legal regime that applies to tracking and data protection.
We concurrently visit the same 743 websites from eight vantage points spanning North America (Ohio, California, Quebec), Europe (France, Germany, Ireland), and Asia (Mumbai, Singapore).
Unless otherwise noted, we focus on Chrome as a high-tracking baseline as shown in~\sectionref{subsec:results-browser}, and we distinguish between \emph{no-click} and \emph{accept-all} consent states.

\begin{figure}[t]
\centering

\setlength{\tabcolsep}{0.45em}

\newcommand{\DonutR}{3}          
\newcommand{\LabelGap}{0.18}     
\tikzset{
  donutlabel/.style={font=\sffamily\bfseries\fontsize{20}{13.5}\selectfont}
}

\begin{tabular}{@{}cccc@{}}
\resizebox{0.2\columnwidth}{!}{%
\begin{tikzpicture}[font=\sffamily, baseline=(current bounding box.center)]
  \wheelchart[data={}, slices style={\WCvarB, draw=black, thick}, radius={2}{3}]
    {38.72/bannerColor,61.28/noBannerColor}
  \wheelchart[data={}, slices style={\WCvarB, draw=black, thick}, radius={0.5}{1.5}]
    {34.59/bannerColor,66.62/noBannerColor}

  \node[donutlabel, anchor=south] at (0,\DonutR+\LabelGap) {Canada};

  \draw (0,0.7)++(right:.2) node[right]{\Large \color{white} \textbf{35\%}};
  \draw (0,2.3)++(right:.2) node[right]{\Large \color{white} \textbf{39\%}};
\end{tikzpicture}} &
\resizebox{0.2\columnwidth}{!}{%
\begin{tikzpicture}[font=\sffamily, baseline=(current bounding box.center)]
  \wheelchart[data={}, slices style={\WCvarB, draw=black, thick}, radius={2}{3}]
    {87.17/bannerColor,12.83/noBannerColor}
  \wheelchart[data={}, slices style={\WCvarB, draw=black, thick}, radius={0.5}{1.5}]
    {60.16/bannerColor,39.84/noBannerColor}

  \node[donutlabel, anchor=south] at (0,\DonutR+\LabelGap) {Ireland};

  \draw (0,0.7)++(right:.2) node[right]{\Large \color{white} \textbf{60\%}};
  \draw (0,2.3)++(right:.2) node[right]{\Large \color{white} \textbf{87\%}};
\end{tikzpicture}} &
\resizebox{0.2\columnwidth}{!}{%
\begin{tikzpicture}[font=\sffamily, baseline=(current bounding box.center)]
  \wheelchart[data={}, slices style={\WCvarB, draw=black, thick}, radius={2}{3}]
    {88.27/bannerColor,11.73/noBannerColor}
  \wheelchart[data={}, slices style={\WCvarB, draw=black, thick}, radius={0.5}{1.5}]
    {60.57/bannerColor,39.43/noBannerColor}

  \node[donutlabel, anchor=south] at (0,\DonutR+\LabelGap) {France};

  \draw (0,0.7)++(right:.2) node[right]{\Large \color{white} \textbf{61\%}};
  \draw (0,2.3)++(right:.2) node[right]{\Large \color{white} \textbf{88\%}};
\end{tikzpicture}} &
\resizebox{0.2\columnwidth}{!}{%
\begin{tikzpicture}[font=\sffamily, baseline=(current bounding box.center)]
  \wheelchart[data={}, slices style={\WCvarB, draw=black, thick}, radius={2}{3}]
    {90.31/bannerColor,9.69/noBannerColor}
  \wheelchart[data={}, slices style={\WCvarB, draw=black, thick}, radius={0.5}{1.5}]
    {59.62/bannerColor,40.38/noBannerColor}

  \node[donutlabel, anchor=south] at (0,\DonutR+\LabelGap) {Germany};

  \draw (0,0.7)++(right:.2) node[right]{\Large \color{white} \textbf{60\%}};
  \draw (0,2.3)++(right:.2) node[right]{\Large \color{white} \textbf{90\%}};
\end{tikzpicture}} \\[1.1cm] 
%
\resizebox{0.2\columnwidth}{!}{%
\begin{tikzpicture}[font=\sffamily, baseline=(current bounding box.center)]
  \wheelchart[data={}, slices style={\WCvarB, draw=black, thick}, radius={2}{3}]
    {27.54/bannerColor,72.46/noBannerColor}
  \wheelchart[data={}, slices style={\WCvarB, draw=black, thick}, radius={0.5}{1.5}]
    {27.86/bannerColor,72.14/noBannerColor}

  \node[donutlabel, anchor=north] at (0,-\DonutR-\LabelGap) {US-California};

  \draw (0,0.7)++(right:.2) node[right]{\Large \color{white} \textbf{28\%}};
  \draw (0,2.3)++(right:.2) node[right]{\Large \color{white} \textbf{28\%}};
\end{tikzpicture}} &
\resizebox{0.2\columnwidth}{!}{%
\begin{tikzpicture}[font=\sffamily, baseline=(current bounding box.center)]
  \wheelchart[data={}, slices style={\WCvarB, draw=black, thick}, radius={2}{3}]
    {21.43/bannerColor,78.57/noBannerColor}
  \wheelchart[data={}, slices style={\WCvarB, draw=black, thick}, radius={0.5}{1.5}]
    {24.36/bannerColor,75.64/noBannerColor}

  \node[donutlabel, anchor=north] at (0,-\DonutR-\LabelGap) {US-Ohio};

  \draw (0,0.7)++(right:.2) node[right]{\Large \color{white} \textbf{24\%}};
  \draw (0,2.3)++(right:.2) node[right]{\Large \color{white} \textbf{21\%}};
\end{tikzpicture}} &
\resizebox{0.2\columnwidth}{!}{%
\begin{tikzpicture}[font=\sffamily, baseline=(current bounding box.center)]
  \wheelchart[data={}, slices style={\WCvarB, draw=black, thick}, radius={2}{3}]
    {26.60/bannerColor,73.4/noBannerColor}
  \wheelchart[data={}, slices style={\WCvarB, draw=black, thick}, radius={0.5}{1.5}]
    {25.03/bannerColor,74.97/noBannerColor}

  \node[donutlabel, anchor=north] at (0,-\DonutR-\LabelGap) {India};

  \draw (0,0.7)++(right:.2) node[right]{\Large \color{white} \textbf{25\%}};
  \draw (0,2.3)++(right:.2) node[right]{\Large \color{white} \textbf{27\%}};
\end{tikzpicture}} &
\resizebox{0.2\columnwidth}{!}{%
\begin{tikzpicture}[font=\sffamily, baseline=(current bounding box.center)]
  \wheelchart[data={}, slices style={\WCvarB, draw=black, thick}, radius={2}{3}]
    {27.95/bannerColor,72.05/noBannerColor}
  \wheelchart[data={}, slices style={\WCvarB, draw=black, thick}, radius={0.5}{1.5}]
    {28.26/bannerColor,71.74/noBannerColor}

  \node[donutlabel, anchor=north] at (0,-\DonutR-\LabelGap) {Singapore};

  \draw (0,0.7)++(right:.2) node[right]{\Large \color{white} \textbf{28\%}};
  \draw (0,2.3)++(right:.2) node[right]{\Large \color{white} \textbf{28\%}};
\end{tikzpicture}} \\
\end{tabular}

\caption{Comparison between the prevalence of websites displaying a banner (inner circle) and their contribution to total \AT\ traffic (outer circle). Red {\textcolor{bannerColor}{$\blacksquare$}} denotes sites with a banner, and blue {\textcolor{noBannerColor}{$\blacksquare$}} denotes sites without.}
\label{figure:results:prevalence-contribution}
\vspace{-.6cm}
\end{figure}
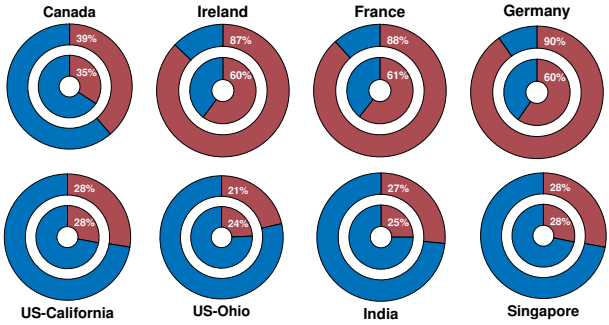

\noindgras{Cookie banner prevalence and tracking contribution.}
\autoref{figure:results:prevalence-contribution} reports, for each vantage point, the fraction of websites displaying a cookie
banner (inner circle) and their contribution to total third-party \AT requests (outer circle).
Banner prevalence varies substantially: EU vantages exhibit the highest rates (60--61\%), consistent with GDPR's consent requirements, while permissive or opt-out locations (Ohio, Mumbai) show significantly fewer banners, with Singapore and Canada (Quebec) falling in between.
This pattern suggests websites selectively deploy consent interfaces based on visitor location.
Critically, in EU, bannered sites contribute the majority of observed tracking, indicating these sites are responsible for most tracking activity.
In non-EU regions, bannered sites contribute a smaller share and more tracking originates from sites without banners.
This suggests consent mechanisms act as gatekeepers for high-intensity tracking deployments, particularly in jurisdictions where consent is legally required.

\begin{figure}[t]
	\centering
	\begin{tikzpicture}
		\begin{axis}[
			width=0.5\columnwidth,
			height=4cm,
			xtick={0,1,2,3,4,5,6,7},
			xticklabels={Canada,France,Germany,India,Ireland,Singapore,US-\\California,US-\\Ohio},
			x tick label style={
				rotate=45,
				anchor=east,
				font=\footnotesize,
				align=right
			},
			xtick=data,
			enlarge x limits={abs=0.4},
			enlarge y limits={upper,value=0.1}, 
			legend style={
				at={(0.8,0.98)}, 
				legend columns=3,
				legend cell align=left,
				fill=white,
				fill opacity=0.9,
				draw=gray!50,
			},
			ybar=0pt,
			bar width=5pt,
			axis x line*=bottom,
			axis y line*=left,
			x=1cm,
			ylabel={Domains},
			ylabel style={
				font=\small,
				yshift=-12pt
			},
			yticklabel style={font=\footnotesize},
			ymajorgrids=true,
			grid style={dashed, gray!25},
			ymin=0,                     
			ymax=65,
			clip=false,                 
			]
			\addplot[fill=beforeclick, draw=none] coordinates {
				(0,36.3) (1,9.3) (2,10) (3,31.2) (4,9.7) (5,33.4) (6,38.7) (7,48.7)
			};
			\addplot[fill=afterclick, draw=none] coordinates {
				(0,48.6) (1,33.9) (2,35.7) (3,37.8) (4,32.3) (5,41.6) (6,46.6) (7,56.2)
			};
			\addplot[fill=violet, draw=none] coordinates {
				(0,45.5) (1,10.1) (2,8.6) (3,37.0) (4,10.4) (5,41.7) (6,46.8) (7,58.4)
			};
			\legend{Before,After,No banner}
		\end{axis}
	\end{tikzpicture}
	\caption{Average third-party \AT domains before/after accepting cookie banners, and from sites with no banner, for Chrome.}
	\label{fig:cookie_banner_impact}
\vspace{-0.6cm}
\end{figure}
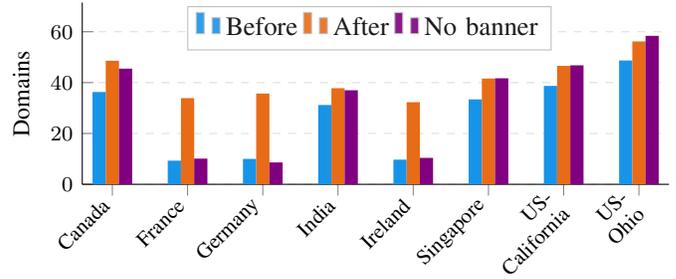

\noindgras{Effect of consent by region.}
We next compare exposure in the ``no-click'' and ``accept-all'' states for each location (\autoref{fig:cookie_banner_impact}).
In GDPR-regulated vantages (FR, DE, IE), baseline tracking in the ``no-click'' state is low: mean \AT counts are significantly smaller than in OH or IN when visiting the same sites without consent.
Once users accept all cookies, third-party \AT exposure in EU locations increases sharply: FR jumps from 9.3 to 33.9 domains (265\% increase).
In contrast, Ohio shows a modest increase from 48.5 to 56.2 domains (16\% increase) because many trackers load even without explicit consent.
This pattern extends to sites without detectable banners: in non-EU regions, these sites contribute substantially higher tracking (37--58 domains) than in EU regions (8--10 domains), further demonstrating how consent requirements shape tracking exposure across the web ecosystem.
These results demonstrate that GDPR-style consent requirements provide a stronger privacy baseline until users click ``Accept'', at which point much of that advantage erodes.
Tracking exposure is also highly skewed: the top 50\% of sites contribute~97\% of all observed \AT domains across all locations (see \autoref{appendix:tracking-distribution} for detailed distribution and category analysis).

\begin{figure*}[t]
  \centering
  \begin{subfigure}[b]{0.45\textwidth}
    \centering
    \includegraphics[width=\textwidth]{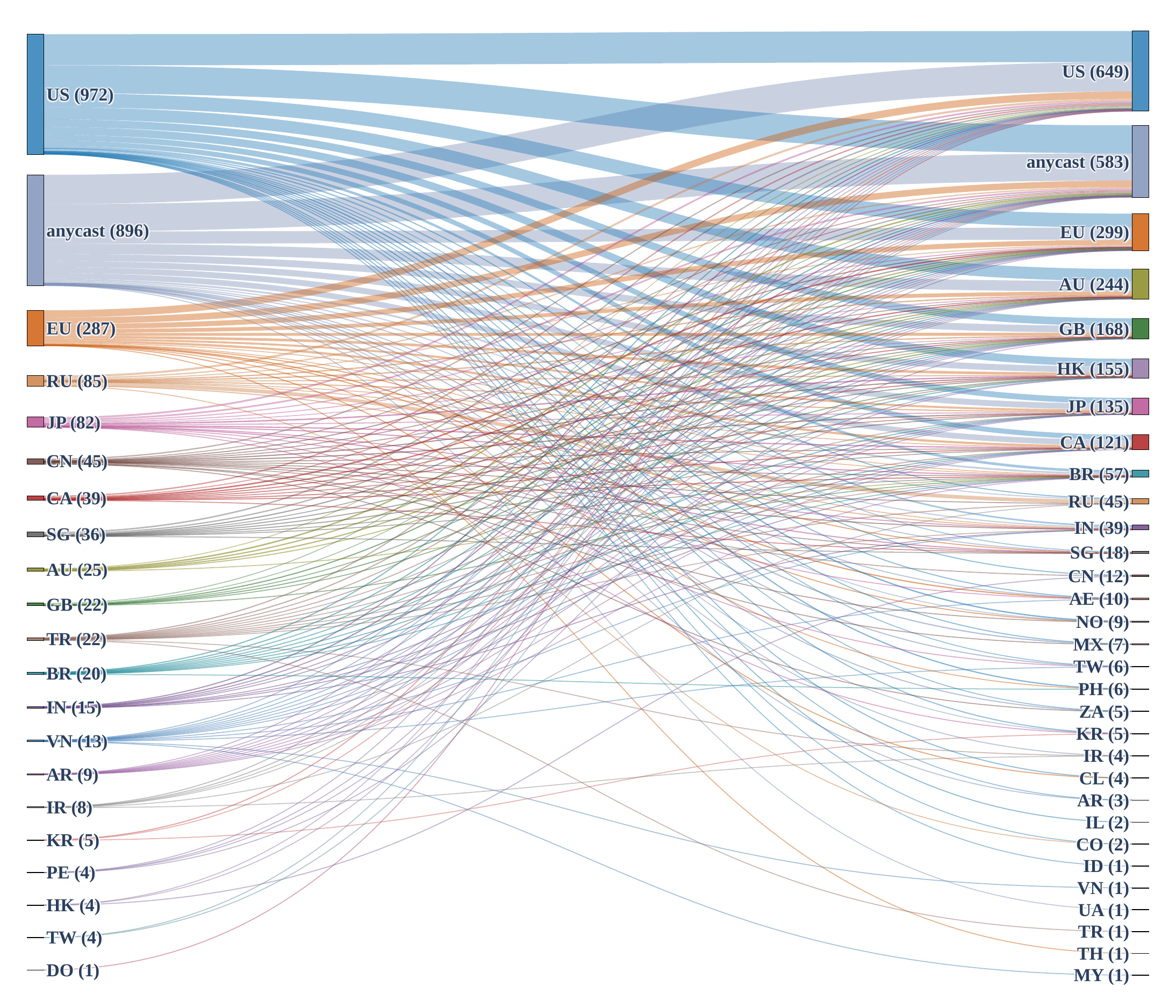}
    \vspace{-.7cm}
    \caption{US-Ohio vantage point (more permissive regulation)}
    \label{fig:sankey_Ohio}
  \end{subfigure}
  \hfill
  \begin{subfigure}[b]{0.45\textwidth}
    \centering
    \includegraphics[width=\textwidth]{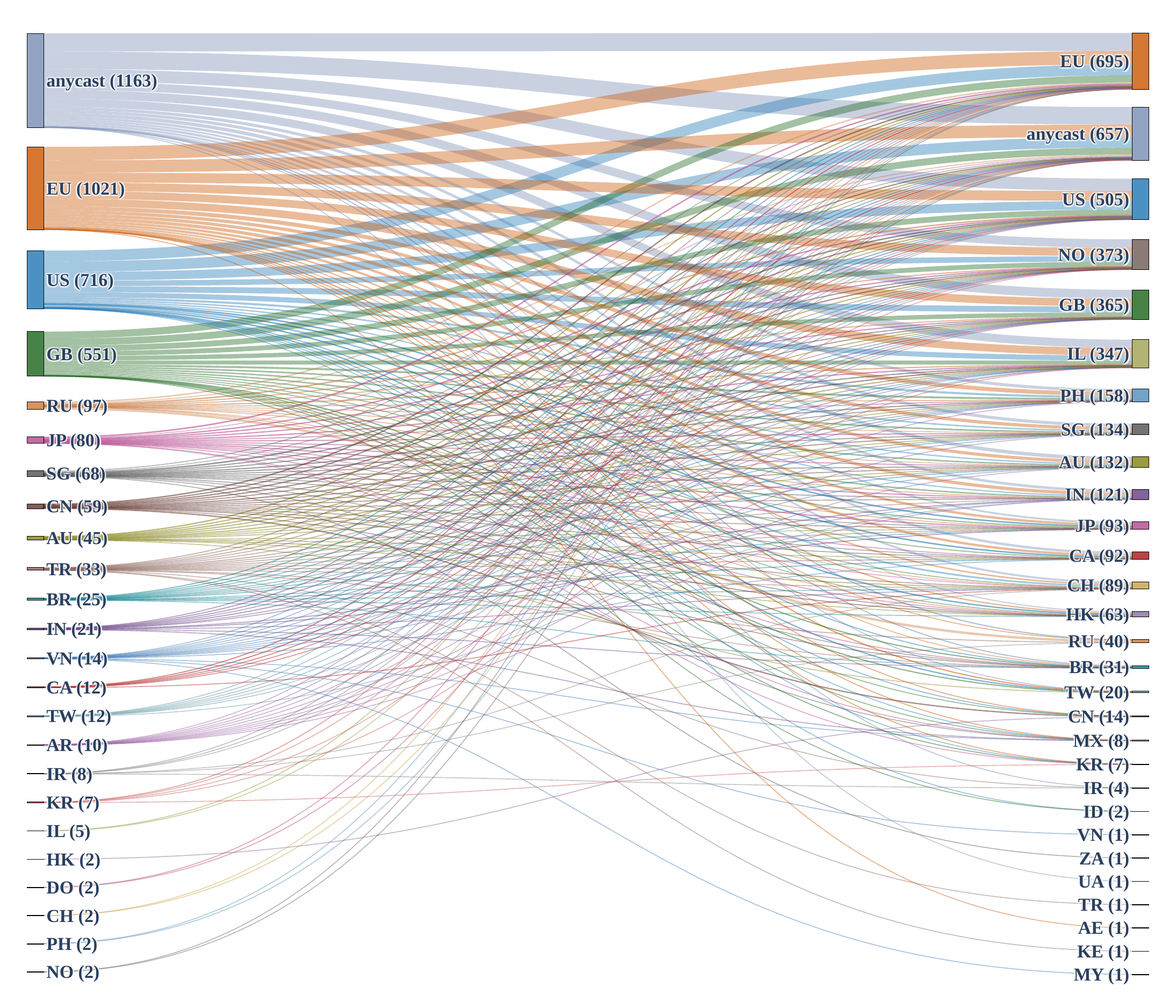}
    \vspace{-.7cm}
    \caption{France vantage point (GDPR regulation)}
    \label{fig:sankey_france}
  \end{subfigure}
  \caption{Cross-border data flows showing first-party (left) and \AT (right) server locations based on IP geo-location.}
  \label{fig:sankey_combined}
  \vspace{-0.5cm}
\end{figure*}
\subsection{Cross-border Data Flows}
\label{subsec:results-dataflow}

Privacy regulations such as GDPR govern not only \emph{what} data may be collected, but also \emph{where} that data may be sent.
For example, GDPR restricts transfers of personal data outside the European Economic Area (EEA) unless the destination country provides an ``adequate level of protection'' or appropriate safeguards are in place~\cite{europaDataProtection}.
Having examined how browser, user location, and hosting jurisdiction affect \AT volumes and cascades, we now ask a complementary question: where do third-party \AT requests go on the network, and how often does \AT traffic leave a user's regulatory region?

\noindgras{IP geolocation.}
Determining server locations is challenging due to Content Delivery Networks (CDNs), anycast routing, and geographically distributed infrastructure.
To obtain robust location estimates, we resolve each third-party apex domain using multiple popular recursive resolvers: Cloudflare (1.1.1.1), Google (8.8.8.8), Quad9 (9.9.9.9), and the default AWS resolver available within our measurement VMs so that we can harvest as many distinct IP addresses as possible for each domain.
Using a diverse set of resolvers allows us to capture location-dependent DNS responses and improves coverage.
Since IP-geolocation can be inaccurate for various reasons~\cite{gill2010dude, Weinberg2018HowTC, darwich2023replication}, we geolocate each resolved IP address using an ensemble of seven GeoIP databases and apply majority voting to select the most likely country for each IP.
This ensemble approach avoids over-reliance on a single database and reduces the impact of individual misclassifications.
Anycast prefixes \cite{githubAnycastprefixesanycatchv4prefixestxtMaster} are filtered where possible to avoid ambiguous locations.
For each (vantage point, domain) pair, we compare the geolocated server country to the user's regulatory region (e.g., EEA versus non-EEA, or the user's own country for non-EU vantages).
If \emph{any} resolver returns an IP outside the user's region, we classify that request as leaving the region.
For EU vantages, we treat the entire EEA as a single regulatory region.

\noindgras{Regional containment of tracking traffic.}
\autoref{tab:server-location} shows that, from every vantage point, third-party \AT requests fan out to a large number of destination countries (45--48 unique countries).
However, the fraction of requests that remain within the user's regulatory region varies markedly across locations.
EU vantages exhibit substantially higher regional containment: France and Germany keep 56\% of requests within the EEA, while Ireland retains 41\%.
In contrast, non-EU vantages show more varied patterns: US vantages retain the highest domestic containment (69--72\%), while India and Singapore show moderate containment (26--31\%), and Canada shows the lowest (6.5\%).
In other words, US users see most tracking processed domestically, EU users see strong regional containment within the EEA, while users in other non-EU regions are considerably more likely to have their data routed to diverse foreign jurisdictions.

To relate this to GDPR's cross-border transfer rules, we further examine whether EU-origin requests are sent to countries that the European Commission has designated as providing ``adequate protection'' under Article~45.
When we count requests that stay within the EEA \emph{or} go to adequacy countries (\autoref{appendix:section-d}), the share of EU requests sent to legally ``adequate'' destinations increases substantially.
This fraction rises to approximately 89\% for FR and IE, and 91\% for DE.
While our GeoIP-based analysis cannot prove legal compliance, it is evident that \AT infrastructure for EU users is preferentially located in jurisdictions that GDPR recognizes as offering comparable protection.
By contrast, traffic from other regions flows more freely to a broader set of destinations, including countries without comprehensive privacy regulations. 

\begin{table}[t]
	\centering
	\caption{Regional containment of third-party \AT requests by vantage point. For FR, DE, and IE, we treat the EEA as a single region, and for OH and CA, the US as a single region. The adequacy column is based on the EU Commission's Adequacy Decision~\cite{europaDataProtection}.}
	\label{tab:server-location}
	\small
	\resizebox{\columnwidth}{!}{%
		\begin{tabular}{lccc}
			\toprule
			\textbf{Vantage point} &
			\textbf{Unique dest.} &
			\textbf{Requests in the} &
			\textbf{Including Adequacy} \\
			& \textbf{countries} & \textbf{same region (\%)} & \textbf{regions (\%)} \\
			\midrule
			Canada           & 45 & \heatmap{6.5}\% & -- \\
			France           & 48 & \heatmap{55}\% & 89\% \\
			Germany          & 47 & \heatmap{56}\% & 91\% \\
			India            & 47 & \heatmap{26}\% & -- \\
			Ireland          & 48 & \heatmap{41}\% & 90\% \\
			Singapore        & 47 & \heatmap{31}\% & -- \\
			USA--California  & 46 & \heatmap{69}\% & -- \\
			USA--Ohio        & 47 & \heatmap{72}\% & -- \\
			\bottomrule
		\end{tabular}
	}
	\vspace{-0.6cm}
\end{table}

\noindgras{First-party vs third-party geographic distribution.}
To better understand these regional patterns, we examine how the geographic locations of first-party website servers differ from those of the third-party \AT requests they trigger.
\autoref{fig:sankey_Ohio} and \autoref{fig:sankey_france} visualize these cross-border data flows for Ohio and France (additional vantage points in \autoref{appendix:sankey}).
Each flow in the Sankey diagram represents the prevalence of a specific cross-border pattern: the flow thickness indicates how many websites in our dataset exhibit that particular combination of first-party server (primary server IP) location and \AT destination region, with each website contributing at most once per unique flow regardless of its tracking volume.
When multiple DNS resolvers return different geographic locations for the same domain, we include all resolved locations, as we cannot definitively determine which location the domain actually serves from.
This prevalence-based view reveals structural patterns in how websites route data across jurisdictions. 
Ohio and France exhibit comparable behavior: first-party servers are concentrated in the EU and US, and most \AT flows terminate in these regions. 
This dominance of the US alongside local infrastructure persists across all vantage points (see \autoref{appendix:sankey}).
\subsection{Hosting Jurisdiction}
\label{subsec:results-host}

A site hosting location is beyond user control but may influence legal obligations and operational choices.
We assign hosting jurisdiction using primary server IP geolocation, restricting analysis to sites with consistent IP-to-country mappings across vantages, after filtering CDNs and anycast~\cite{githubAnycastprefixesanycatchv4prefixestxtMaster}.
\autoref{fig:host-combined} shows cookie banner prevalence vs.\ \AT contribution by hosting jurisdiction, aggregated by EU (blue) or non-EU (red) user location (detailed breakdowns in \autoref{appendix:host-location}).
We identify four patterns:
\begin{figure}[tb]
    \centering
	\resizebox{\columnwidth}{!}{
    \begin{tikzpicture}
    	\begin{axis}[
    		width=\columnwidth,
    		height=6cm,
    		xlabel={Cookie Banner Prevalence (\%)},
    		ylabel={Ad \& Tracking Contribution (\%)},
    		grid=major,
    		xmin=0, xmax=100,
    		ymin=0, ymax=100,
    		]
    		\addplot[only marks, mark=*, mark size=3pt, blue, draw=black, line width=0.5pt, opacity=0.5] coordinates {
    			(62.4,81.8) (60.6,96.6) (33.3,89.7) (50,81.1) (50,80.8) (38.4,67.4) (18.75,78.3)
    		};
    		\node[blue] at (axis cs:60.4,81.8) [anchor=south west] {\scriptsize US};
    		\node[blue] at (axis cs:60.6,96.6) [anchor=north] {\scriptsize EU};
    		\node[blue] at (axis cs:33.3,89.7) [anchor=west] {\scriptsize SG};
    		\node[blue] at (axis cs:50,81.1) [anchor=south east] {\scriptsize IN};
    		\node[blue] at (axis cs:50,80.8) [anchor=north west] {\scriptsize CA};
    		\node[blue] at (axis cs:48.4,67.4) [anchor=east] {\scriptsize RU};
    		\node[blue] at (axis cs:18.75,78.3) [anchor=south east] {\scriptsize CN};

    		\addplot[only marks, mark=square*, mark size=3pt, red, draw=black, line width=0.5pt, opacity=0.5] coordinates {
    			(32.9,19.0) (42.5,56.7) (11.1,1.0) (25,6.4) (0,0) (25.5,33.7) (18.75,77.4)
    		};
    		\node[red] at (axis cs:32.9,19.0) [anchor=south] {\scriptsize US};
    		\node[red] at (axis cs:42.5,56.7) [anchor=west] {\scriptsize EU};
    		\node[red] at (axis cs:11.1,6) [anchor=west] {\scriptsize SG};
    		\node[red] at (axis cs:25,6.4) [anchor=west] {\scriptsize IN};
    		\node[red] at (axis cs:8,8) [anchor=north east] {\scriptsize CA};
    		\node[red] at (axis cs:25.5,33.7) [anchor=east] {\scriptsize RU};
    		\node[red] at (axis cs:18.75,77.4) [anchor=north west] {\scriptsize CN};

    		\draw[thick, black, dashed]
    		(axis cs:18.75,77.85) ellipse [x radius=8, y radius=10];
    		\node[font=\scriptsize] at (axis cs:23,83) {\textbf{D}};

    		\draw[thick, black, dashed]
    		(axis cs:18,16) ellipse [x radius=22, y radius=30];
    		\node[font=\scriptsize] at (axis cs:10,28) {\textbf{A}};

    		\draw[thick, black, dashed]
    		(axis cs:48,84) ellipse [x radius=20, y radius=24];
    		\node[font=\scriptsize] at (axis cs:58,70) {\textbf{B}};

    		\draw[thick, black, dashed]
    		(axis cs:42.5,56.7) ellipse [x radius=7, y radius=4.5];
    		\node[font=\scriptsize] at (axis cs:38,57) {\textbf{C}};
    	\end{axis}
    \end{tikzpicture}
	}
    \caption{Nodes represents the hosting locations (\eg CA, CN, EU, \etc).
    Blue circles ({\textcolor{blue}{$\bullet$}}) are hosts accessed from EU whereas red squares ({\textcolor{red}{$\blacksquare$}}) are accessed by non-EU vantages.
    The x-axis shows the prevalence of cookie banners, the y-axis shows the contribution of websites with cookie banners to the total \AT in their location.
    We identify 4 regions in the figure (A, B, C, and D).
    (More detailed data in appendix \autoref{appendix:host-location}.)
    }
    \label{fig:host-combined}
	\vspace{-0.5cm}
    \end{figure}
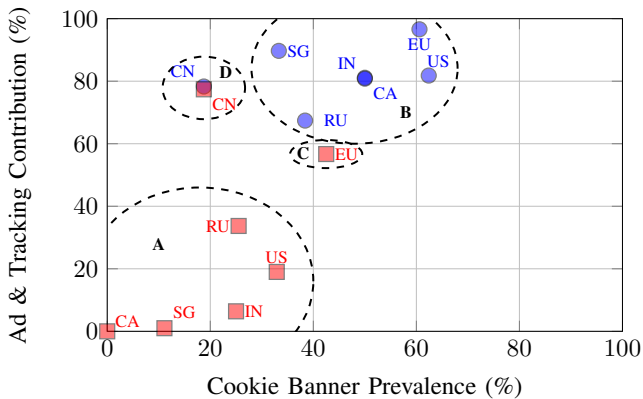

\noindent \emph{A/ Non-EU baseline:} Non-EU users accessing sites hosted outside CN/EU see infrequent banners with low \AT contribution from bannered sites, reflecting permissive environments where consent is optional.

\noindent \emph{B/ EU adaptation:} Regardless of hosting jurisdiction, sites present banners frequently to EU users (61--64\% for US-hosted sites) with bannered sites contributing 81--83\% of \AT, demonstrating location-based tailoring for GDPR compliance.

\noindent \emph{C/ EU-hosted baseline:} EU-hosted sites display banners at moderate rates (42--44\%) even to non-EU users, suggesting GDPR-compliant sites deploy consent mechanisms uniformly.

\noindent \emph{D/ CN hosts:} CN-hosted sites show uniform 19\% banner prevalence regardless of user location, consistent with China's PIPL~\cite{chinaprivacylaw} mandating GDPR-like consent applied globally.

\section{Discussion}
\label{sec:discussion}
Our results show browser choice, browsing location and its regulatory framework, and hosting jurisdiction each affect different aspects of the \AT ecosystem: baseline volume, cookie-consent behavior, cascade structure, and cross-border data flows.
We discuss practical implications and limitations.

\subsection{Implications}

Privacy-focused browsers reduce \AT exposure by 30\% in permissive regions but offer smaller gains where GDPR already lowers baseline \AT.
Accepting cookies erodes this advantage: in FR, \AT domains per site increase~3x after consent vs. smaller increases in OH.
GDPR-regulated locations exhibit 80\% fewer \AT domains pre-consent and keep 90\% of requests within EEA/adequacy countries, suggesting opt-in consent and cross-border rules meaningfully reduce tracking and contain data flows.
However, dark patterns steering users toward ``Accept all'' can undo much of this benefit.
Location-based consent discrimination is widespread and easily measurable: websites present banners 2--3x more frequently to EU visitors, patterns amenable to regulatory scrutiny.

\subsection{Limitations}
\label{disc:limitation}
As with any empirical study, our work involves trade-offs limiting generalizability.
We focus on 1{,}005 popular websites and 8 vantages in North America, Europe, and Asia, potentially missing smaller or localized sites and some countries within each region.
Our \AT blocklists (e.g., EasyList/EasyPrivacy~\cite{easylist-default,Easylist-privacy}) are designed for URL-level ad blocking; applying them at the domain level may over-count (flagging benign requests to listed domains) or under-count (missing path-specific first-party tracking rules).
We mitigate this using Wally3K's curated lists~\cite{wally3k,firebog-site}, which filter entries unsuitable for domain-level matching, though some error may remain.
This also explains why \AT domains appear in Brave crawls: Brave blocks specific URL patterns, while our classification flags any request to a listed domain.
Crucially, comparative analysis across browsers and regions remains valid as the methodology is applied uniformly.
Our lexicon-driven cookie-banner detection has high but imperfect coverage, and we model only two consent states; we measure tracking behavior rather than legal compliance.
Our network-level observations may miss CNAME cloaking~\cite{paloaltonetworksCNAMECloaking} and do not account for data volume or sensitivity; GeoIP-based cross-border analysis may contain errors from CDNs and anycast.
Our factorial design observes associations rather than proving causal effects; site mix, business models, and regional ad markets may contribute.
Finally, our setup cost \$1K+ USD, generated 1.5~TB data, and required 2.5~weeks, limiting temporal repetition and additional dimensions (e.g., mobile browsers).

\section{Related Work}
\label{sec:rw}
We focus on recent work that evaluates tracking under regulatory constraints, incorporates geographic scope, or foregrounds measurement methodology.

Studies operationalize GDPR/CCPA requirements through measurements of user-facing behavior:
Sørensen et al.~\cite{sorensen2019before} quantify third-party presence before/after GDPR;
Sanchez-Rola et al.~\cite{sanchez2019can} examine tracking persistence after opt-out;
Liu et al.~\cite{liu2024opted} audit consent choices across GDPR/CCPA contexts;
Hausladen et al.~\cite{hausladen2025websites} evaluate GPC signal compliance.
Iordanou et al.~\cite{iordanou2018tracing} characterize cross-border tracking endpoints for EU users;
Vallina et al.~\cite{vallina2019tales} study tracking across multiple vantages;
Singh et al.~\cite{SinghIMC25} broaden coverage to 23 Global South countries.
Urban et al.~\cite{urban2020beyond} measure third-party dynamics beyond landing pages;
Stafeev et al.~\cite{stafeev2024sok} systematize crawling design space;
Hantke et al.~\cite{hantke2025web} propose reproducible measurement tooling.
\sysname builds on these insights by jointly varying browser, user location, hosting jurisdiction, and consent state within a controlled factorial design.
A chronological summary table of related work in comparison to \sysname is in Appendix \autoref{app:rw}.

\section{Conclusion}
\label{sec:conclusion}
We examined how third-party \AT exposure varies across browser, location, and hosting jurisdiction through synchronized,
consent-aware measurements of 743 sites across 8 vantage points, 4 browsers, and 2 consent states.
Browsing location is the strongest predictor, influencing pre-consent baselines, consent interface prevalence, and post-consent \AT levels.
Browser choice provides context-dependent leverage, with larger gains in permissive settings.
Hosting jurisdiction is weaker, suggesting sites adapt to inferred user location rather than hosting location.
EU vantages show higher regional containment of \AT traffic, especially to EEA and adequacy destinations.
Our results show user-controllable choices matter, but structural context, location-conditioned consent gating and region-specific infrastructure--often dominates,
providing measurable compliance signals for regulators and emphasizing the need to treat browser, location, and consent state as first-class experimental variables.

\section*{Acknowledgment}
\noindent We are grateful to the anonymous reviewers for their constructive feedback and guidance in shaping this paper.
We acknowledge the support of the Natural Sciences and Engineering Research Council of Canada (NSERC).
Nous remercions le Conseil de recherches en sciences naturelles et en génie du Canada (CRSNG) de son soutien.
We acknowledge the support of the Canadian Foundation  for Innovation (CFI).
Nous remercions la Fondation canadienne pour l’innovation (FCI) de son soutien.
Finally, we acknowledge the support of the British Columbia Knowledge Development Fund (BCKDF), UBC Advanced Research Computing (ARC), and Dell Computer.
Any opinions, findings, and conclusions or recommendations expressed in this material are those of the authors and do not reflect those of the sponsors.

\balance  
\bibliographystyle{IEEEtran}
\bibliography{biblio}

\appendix
\subsection{Data and Code Availability}
\label{appendix:section-a}

To support reproducibility and future research, we publicly release our dataset (raw HAR files, processed tracking-domain classifications, and aggregated statistics) and the \sysname source code, analysis pipelines, and visualization scripts at \url{https://github.com/ubc-spg/RegTrack}.

\subsection{Use of Generative AI}
\label{appendix:section-b}

We used generative AI tools (ChatGPT, Claude) to assist with specific technical and editorial tasks during this research, and we document their use here for transparency.
For writing, we relied on these tools to improve sentence clarity, correct grammatical errors, rephrase awkward constructions to enhance readability, and help maintain consistent terminology throughout the paper.
For code, we used them to generate matplotlib plotting routines for data visualizations and to draft wrapper functions for data processing pipelines.
All AI-generated code and text were reviewed, validated, and, where necessary, modified by the authors to ensure accuracy and appropriateness.
Finally, we used AI tools to automatically classify invalid pages, with more details provided in the next subsection (\autoref{appendix:section-c}).

\subsection{Finding Invalid Pages using Llama4}
\label{appendix:section-c}

We use vision-language models (VLMs) to automatically identify invalid pages in our crawl data.
Invalid pages include CAPTCHAs, error pages, security warnings, and connection failures that would skew our tracking measurements.
To select the best model, we evaluated five VLMs against 400 manually labeled screenshots (\autoref{tab:model_comparison}).
Llama4 achieves the highest accuracy (99\%) and Cohen's $\kappa$ (0.94), indicating near-perfect agreement with human labels.
\autoref{figure:prompt} shows the prompt used to classify each captured page.
We exclude websites whose landing page is classified as invalid (result = 1) in more than 50\% of visits in any browser-location configuration, ensuring that our tracking measurements reflect actual website behavior rather than error states.

\begin{table}[H]
	\centering
	\caption{Performance of vision-language models for webpage screenshot classification (success vs.\ failure).}
	\label{tab:model_comparison}
	\begin{tabular}{l*{5}{c}}
	\toprule
	\textbf{Model} & \textbf{Acc.} & \textbf{Prec.} & \textbf{Rec.} & \textbf{F1} & \textbf{Kappa} \\
	\midrule
	llama4          & \heatmaps{0.99} & \heatmaps{0.98} & \heatmaps{0.93} & \heatmaps{0.95} & \heatmaps{0.94} \\
	Qwen2.5 VL      & \heatmaps{0.98} & \heatmaps{0.95} & \heatmaps{0.93} & \heatmaps{0.94} & \heatmaps{0.93} \\	llama3.2-vision & \heatmaps{0.91} & \heatmaps{0.92} & \heatmaps{0.52} & \heatmaps{0.67} & \heatmaps{0.62} \\
	LLaVA 7B        & \heatmaps{0.91} & \heatmaps{0.76} & \heatmaps{0.63} & \heatmaps{0.69} & \heatmaps{0.63} \\
	Gemma3          & \heatmaps{0.32} & \heatmaps{0.19} & \heatmaps{0.97} & \heatmaps{0.32} & \heatmaps{0.06} \\
	\bottomrule
	\end{tabular}
\end{table}

\begin{figure}[H]
\centering
\begin{lstlisting}[
  basicstyle=\small\ttfamily,
  breaklines=true,
  columns=fullflexible,
  frame=single,              % Adds a box around the listing
  frameround=tttt,           % Optional: rounded corners (tttt = all corners)
  framesep=3pt,             % Space between frame and text
  backgroundcolor=\color{gray!5},  % Optional: light gray background
]
Does this webpage show ANY of these invalid page indicators:
- CAPTCHA verification or "I'm not a robot" checkbox
- "Please verify you are human" messages
- Security warnings or "Potential Security Risk"
- Connection errors like "This site can't be reached" or "can't reach this page"
- DNS errors or technical error codes
- 404/403/500 error messages or "Not Found"
- Generic error messages like "Something went wrong" or "We're having trouble"
- "Unable to connect" or connection timeout messages
- Blank or mostly empty pages with minimal content
- Browser error pages or access restrictions
- "Access Denied" or permission error messages
- Security warnings like "This site has been reported as unsafe"
- Technical service pages showing raw data or configuration
- Completely blank white/empty pages with no content
- Microsoft Defender or browser security warnings

Answer in JSON format:
{
  "result": 1 or 0 (1 for YES, 0 for NO),
  "reason": "brief explanation in 30 words or less"
}
\end{lstlisting}
\caption{LLM prompt for invalid page detection.}
\label{figure:prompt}
\end{figure}

\subsection{Blocklists Used for A\&T Classification}
\label{appendix:blocklists}

\autoref{tab:blocklists} lists the blocklists used to classify third-party domains as advertising and tracking (\AT) as well as their descriptions.

\begin{table}[H]
	\caption{Blocklists used for \AT identification.}
	\label{tab:blocklists}
	\centering
	\footnotesize
	\setlength{\extrarowheight}{2pt}
	\begin{tabular}{@{}lp{0.62\columnwidth}@{}}
		\toprule
		\textbf{Blocklist} & \textbf{Description} \\
		\midrule
		AdGuard DNS~\cite{adguard-default}
		& Default blocklist for AdGuard DNS service \\
		\hdashline[0.5pt/1pt]

		StevenBlack~\cite{steven-black}
		& Default blocklist for Pi-hole ad blocking \\
		\hdashline[0.5pt/1pt]

		LanikSJ Admiral~\cite{admiral_firebog}
		& Blocks ad-blocker detectors like Admiral \\
		\hdashline[0.5pt/1pt]

		Anudeep AdServers~\cite{anudeep}
		& Ad server list by AnudeepND (via NextDNS) \\
		\hdashline[0.5pt/1pt]

		EasyList Default~\cite{easylist-default}
		& Primary ad blocking list for AdBlock, AdGuard, uBlock Origin \\
		\hdashline[0.5pt/1pt]

		EasyList Privacy~\cite{Easylist-privacy}
		& Tracker blocking companion to EasyList \\
		\hdashline[0.5pt/1pt]

		Firebog Prigent~\cite{firebog-prigent}
		& Ad list by Fabrice Prigent \\
		\hdashline[0.5pt/1pt]

		Frogeye~\cite{frogeye-first}
		& Tracker list by Geoffrey Frogeye \\
		\bottomrule
	\end{tabular}
\end{table}

\subsection{GDPR Adequacy Destinations}
\label{appendix:section-d}

Under GDPR Article 45, the European Commission has issued adequacy decisions~\cite{europaDataProtection} for the following countries and territories, allowing personal data to flow from the EU to them without additional safeguards: Andorra, Argentina, Canada (commercial organizations), the Faroe Islands, Guernsey, Israel, the Isle of Man, Japan, Jersey, New Zealand, the Republic of Korea, Switzerland, the United Kingdom, Uruguay, and the United States of America (under the EU--U.S. Data Privacy Framework).

In our server-IP analysis (\autoref{subsec:results-dataflow}), we therefore treat data transfers to these jurisdictions as compliant with the GDPR's cross-border transfer restrictions.

\subsection{Tracking Distribution and Category Analysis}
\label{appendix:tracking-distribution}

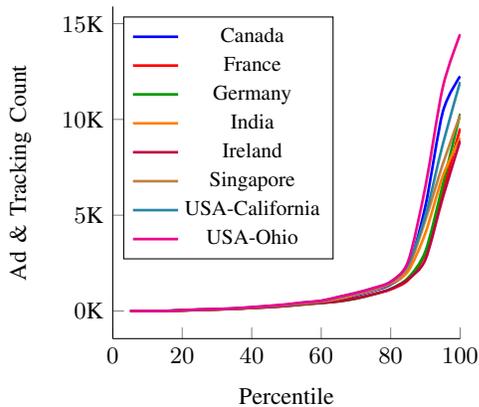
\begin{figure}[t]
  \centering
  \begin{tikzpicture}
  	\begin{axis}[
  		width=0.7\columnwidth,
  		height=6cm,
  		xmin=0, xmax=100,
  		xlabel={Percentile},
  		ylabel={Ad \& Tracking Count},
  		axis x line* = bottom,
  		axis y line* = left,
  		legend pos=north west,
  		legend style={font=\footnotesize}, 
  		scaled y ticks=false,
  		scaled x ticks=false,
  		yticklabel={\pgfmathparse{\tick/1000}\pgfmathprintnumber{\pgfmathresult}K},
  		xticklabel style={/pgf/number format/fixed},
  		every axis plot/.append style={line width=1pt}, 
  		label style={font=\small},  
  		tick label style={font=\small}, 
  		]
  		
  		\addplot[smooth, mark=none, color=blue]
  		coordinates {
  			(5,0)
  			(10,0)
  			(15,0)
  			(20,36.078)
  			(25,72.156)
  			(30,93.8028)
  			(35,119.0574)
  			(40,169.5666)
  			(45,230.8992)
  			(50,295.8396)
  			(55,389.6424)
  			(60,469.014)
  			(65,591.6792)
  			(70,815.3628)
  			(75,1057.0854)
  			(80,1385.3952)
  			(85,2226.0126)
  			(90,5483.856)
  			(95,10390.464)
  			(100,12252.0888)
  		};
  		
  		\addplot[smooth, mark=none, color=red]
  		coordinates {
  			(5,0)
  			(10,0)
  			(15,0)
  			(20,20.1352)
  			(25,52.8549)
  			(30,80.5408)
  			(35,110.7436)
  			(40,156.0478)
  			(45,198.8351)
  			(50,259.2407)
  			(55,344.8153)
  			(60,422.8392)
  			(65,503.38)
  			(70,644.3264)
  			(75,860.7798)
  			(80,1127.5712)
  			(85,1613.3329)
  			(90,3020.28)
  			(95,6221.7768)
  			(100,9531.5003)
  		};
  		
  		\addplot[smooth, mark=none, color=green!60!black]
  		coordinates {
  			(5,0)
  			(10,0)
  			(15,0)
  			(20,21.2344)
  			(25,55.7403)
  			(30,76.9747)
  			(35,111.4806)
  			(40,151.2951)
  			(45,201.7268)
  			(50,262.7757)
  			(55,339.7504)
  			(60,416.7251)
  			(65,499.0084)
  			(70,647.6492)
  			(75,870.6104)
  			(80,1159.9291)
  			(85,1725.295)
  			(90,3118.8025)
  			(95,6587.9726)
  			(100,10296.0297)
  		};
  		
  		\addplot[smooth, mark=none, color=orange]
  		coordinates {
  			(5,0)
  			(10,0)
  			(15,0)
  			(20,28.075)
  			(25,70.1875)
  			(30,87.0325)
  			(35,120.7225)
  			(40,168.45)
  			(45,238.6375)
  			(50,294.7875)
  			(55,395.8575)
  			(60,482.89)
  			(65,609.2275)
  			(70,853.48)
  			(75,1086.5025)
  			(80,1426.21)
  			(85,2069.1275)
  			(90,4068.0675)
  			(95,7038.4025)
  			(100,9037.3425)
  		};
  		
  		\addplot[smooth, mark=none, color=purple]
  		coordinates {
  			(5,0)
  			(10,0)
  			(15,0)
  			(20,24.015)
  			(25,60.0375)
  			(30,76.848)
  			(35,112.8705)
  			(40,156.0975)
  			(45,204.1275)
  			(50,266.5665)
  			(55,357.8235)
  			(60,425.0655)
  			(65,509.118)
  			(70,667.617)
  			(75,878.949)
  			(80,1138.311)
  			(85,1647.429)
  			(90,2687.2785)
  			(95,5917.296)
  			(100,8885.55)
  		};
  		
  		\addplot[smooth, mark=none, color=brown]
  		coordinates {
  			(5,0)
  			(10,0)
  			(15,0)
  			(20,30.878)
  			(25,71.0194)
  			(30,92.634)
  			(35,123.512)
  			(40,176.0046)
  			(45,240.8484)
  			(50,305.6922)
  			(55,395.2384)
  			(60,481.6968)
  			(65,636.0868)
  			(70,870.7596)
  			(75,1102.3446)
  			(80,1475.9684)
  			(85,2244.8306)
  			(90,4745.9486)
  			(95,7667.0074)
  			(100,10217.5302)
  		};
  		
  		\addplot[smooth, mark=none, color=cyan!60!black]
  		coordinates {
  			(5,0)
  			(10,0)
  			(15,0)
  			(20,44.9748)
  			(25,76.1112)
  			(30,100.3284)
  			(35,134.9244)
  			(40,197.1972)
  			(45,262.9296)
  			(50,332.1216)
  			(55,442.8288)
  			(60,532.7784)
  			(65,733.4352)
  			(70,944.4708)
  			(75,1186.6428)
  			(80,1515.3048)
  			(85,2449.3968)
  			(90,4985.2836)
  			(95,8700.894)
  			(100,11956.3776)
  		};
  		
  		\addplot[smooth, mark=none, color=magenta]
  		coordinates {
  			(5,0)
  			(10,0)
  			(15,0)
  			(20,45.9349)
  			(25,70.9903)
  			(30,100.2216)
  			(35,137.8047)
  			(40,200.4432)
  			(45,267.2576)
  			(50,346.5997)
  			(55,450.9972)
  			(60,530.3393)
  			(65,734.9584)
  			(70,939.5775)
  			(75,1194.3074)
  			(80,1532.5553)
  			(85,2580.7062)
  			(90,6497.7004)
  			(95,11679.9923)
  			(100,14448.614)
  		};
  		
  		\legend{
  			Canada,
  			France,
  			Germany,
  			India,
  			Ireland,
  			Singapore,
  			USA-California,
  			USA-Ohio
  		}
  		
  	\end{axis}
  \end{tikzpicture}
  \caption{CDF of 3rd-party \AT domains across website percentiles (5-point increments) by location. X=100 represents the 95th-100th percentile. The steep right-to-left decline demonstrates that 
  \AT is concentrated in higher percentile websites.}
  \label{fig:distribution_percentile}
\end{figure}

\noindgras{Skewed distribution of tracking across sites.}
Tracking exposure is highly skewed across sites: most sites contact relatively few third-party \AT domains, while a small fraction contact dozens or even hundreds.
\autoref{fig:distribution_percentile} illustrates this concentration by showing the distribution of third-party \AT domains across website percentiles.
The top 50\% of sites contribute roughly 97\% of all observed third-party \AT apex domains in our Chrome accept-all configuration.
This pattern holds consistently across all vantage points.

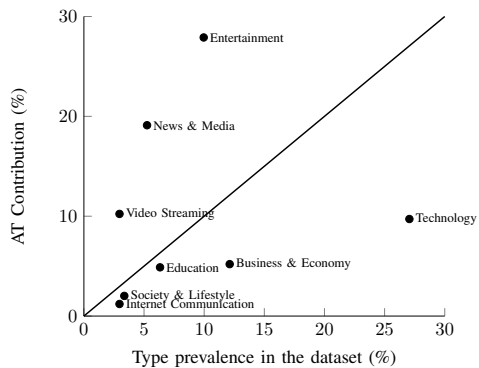
\begin{figure}[H]
  \centering
  \def\a{1}
  \begin{tikzpicture}[scale=0.7]
      \begin{axis}[
          xmin=0, xmax=30,
          ymin=0, ymax=30,
          xlabel={Type prevalence in the dataset (\%)},
          ylabel={AT Contribution (\%)},
          axis x line* = bottom,
          axis y line* = left,
          clip=false,
          ]
          \addplot[thick, domain=0:30, samples=2] {\a*x};
          \addplot[
          only marks,
          mark=*,
          nodes near coords,
          point meta=explicit symbolic,
          every node near coord/.append style={font=\scriptsize, anchor=west},
          ]
          table[x=prevalence, y=contrib, meta=label] {
              label                     prevalence  contrib
              Technology                27.06       9.72
              {Business \& Economy}     12.12       5.2
              Entertainment             9.96        27.9
              Education                 6.33        4.88
              {News \& Media}           5.25        19.11
              {Society \& Lifestyle}    3.36        2.02
              {Video Streaming}         2.96        10.23
              {Internet Communication}  2.96        1.21
          };
      \end{axis}
  \end{tikzpicture}
  \caption{Category prevalence vs. \AT contribution in US-Ohio using Chrome. Categories above the diagonal contribute disproportionately to \AT.}
  \label{fig:category}
\end{figure}

\noindgras{Which sites track the most?}
To understand which website types contribute more to \AT, we group sites by category.
\autoref{fig:category} shows, for Ohio with Chrome, the relationship between each category's prevalence in our dataset and its contribution to total \AT requests.
Categories above the diagonal contribute disproportionately high tracking: Entertainment (10\% of sites, 28\% of \AT), News \& Media (5\% of sites, 19\% of \AT), and Video Streaming (3\% of sites, 10\% of \AT).

\subsection{Measurement Infrastructure}
\label{appendix:section-g}

Our measurement infrastructure consists of distributed crawling nodes and a centralized orchestration layer.

\noindgras{Regional Crawling Nodes.}
We deploy AWS EC2 instances in eight geographic regions to perform web crawls:
\begin{itemize}[leftmargin=*]
	\item \textbf{North America:} Ohio (us-east-2), California (us-west-1), Canada (ca-central-1)
	\item \textbf{Europe:} Ireland (eu-west-1), Germany (eu-central-1), France (eu-west-3)
	\item \textbf{Asia:} Singapore (ap-southeast-1), India (ap-south-1)
\end{itemize}

All crawling nodes use the same instance type to ensure measurement consistency with the following specifications:
\begin{itemize}[leftmargin=*]
	\item \textbf{Instance type:} m6a.32xlarge
	\item \textbf{CPU:} 128 vCPUs
	\item \textbf{Memory:} 512 GB RAM
	\item \textbf{Network:} 50 Gbps
	\item \textbf{Disk:} 300 GB gp3 SSD with 20k IOPS and 1 GB/s bandwidth
	\item \textbf{OS:} Ubuntu 22.04 LTS
\end{itemize}

Each node runs Browsertime to automate browser interactions and collect HAR files that record all network requests.

\noindgras{Central Orchestration Server.}
We use a Dell PowerEdge R750 server to coordinate crawls across all regions and process the collected data:
\begin{itemize}[leftmargin=*]
	\item \textbf{Model:} Dell PowerEdge R750
	\item \textbf{CPU:} 64 cores (2x Intel Xeon Gold 6326 @ 2.90 GHz)
	\item \textbf{Memory:} 1024 GB RAM
\end{itemize}

This server schedules crawls, monitors progress across regions, collects HAR files from the crawling nodes, and performs initial aggregation and full analysis.

\noindgras{Invalid Page Classification Server.}
We use a GPU-equipped Dell PowerEdge R750 server for Llama4-based invalid-page classification:
\begin{itemize}[leftmargin=*]
	\item \textbf{Model:} Dell PowerEdge R750
	\item \textbf{CPU:} 64 cores (2x Intel Xeon Gold 6326 @ 2.90 GHz)
	\item \textbf{Memory:} 1024 GB RAM
	\item \textbf{GPU:} NVIDIA A100 PCIe 80 GB
\end{itemize}

\subsection{Additional Cross-Border Data Flow Diagrams}
\label{appendix:sankey}

\begin{figure*}[p]
	\centering
	\begin{subfigure}[b]{0.48\textwidth}
		\centering
		\includegraphics[width=\textwidth]{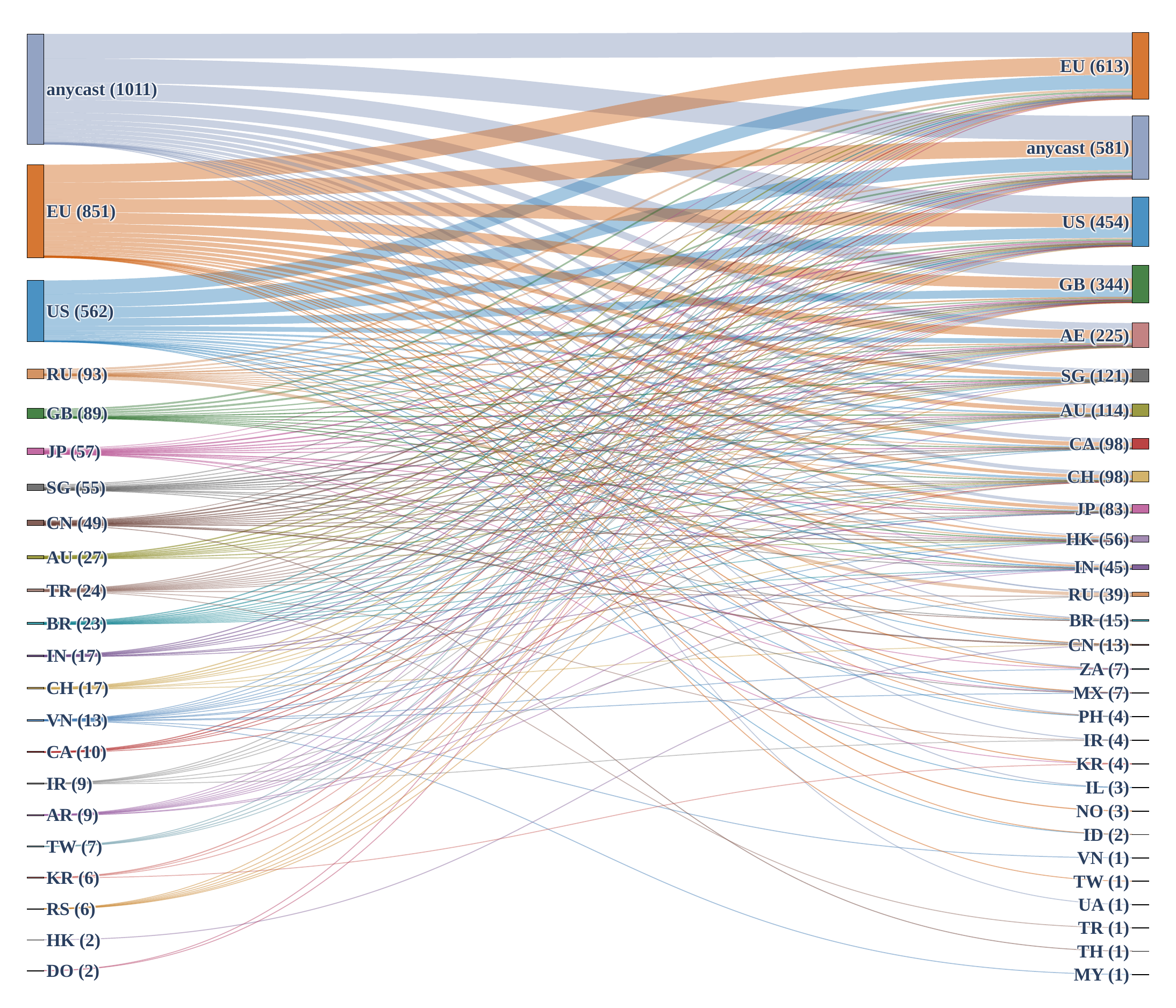}
		\vspace{-1cm}
		\caption{Germany}
		\label{fig:sankey_Germany}
	\end{subfigure}
	\begin{subfigure}[b]{0.48\textwidth}
		\centering
		\includegraphics[width=\textwidth]{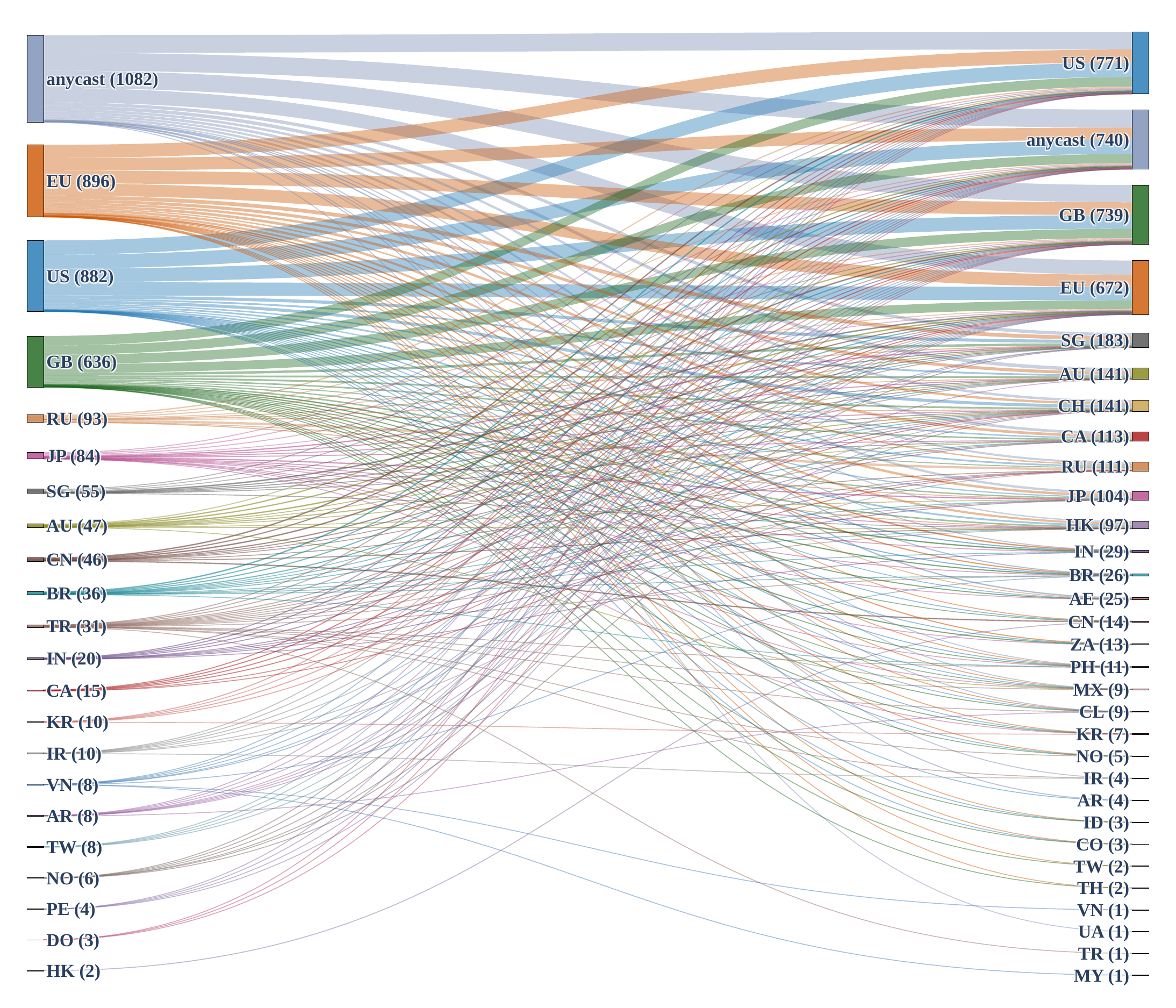}
		\vspace{-1cm}
		\caption{Ireland}
		\label{fig:sankey_Ireland}
	\end{subfigure}

	\begin{subfigure}[b]{0.48\textwidth}
		\centering
		\includegraphics[width=\textwidth]{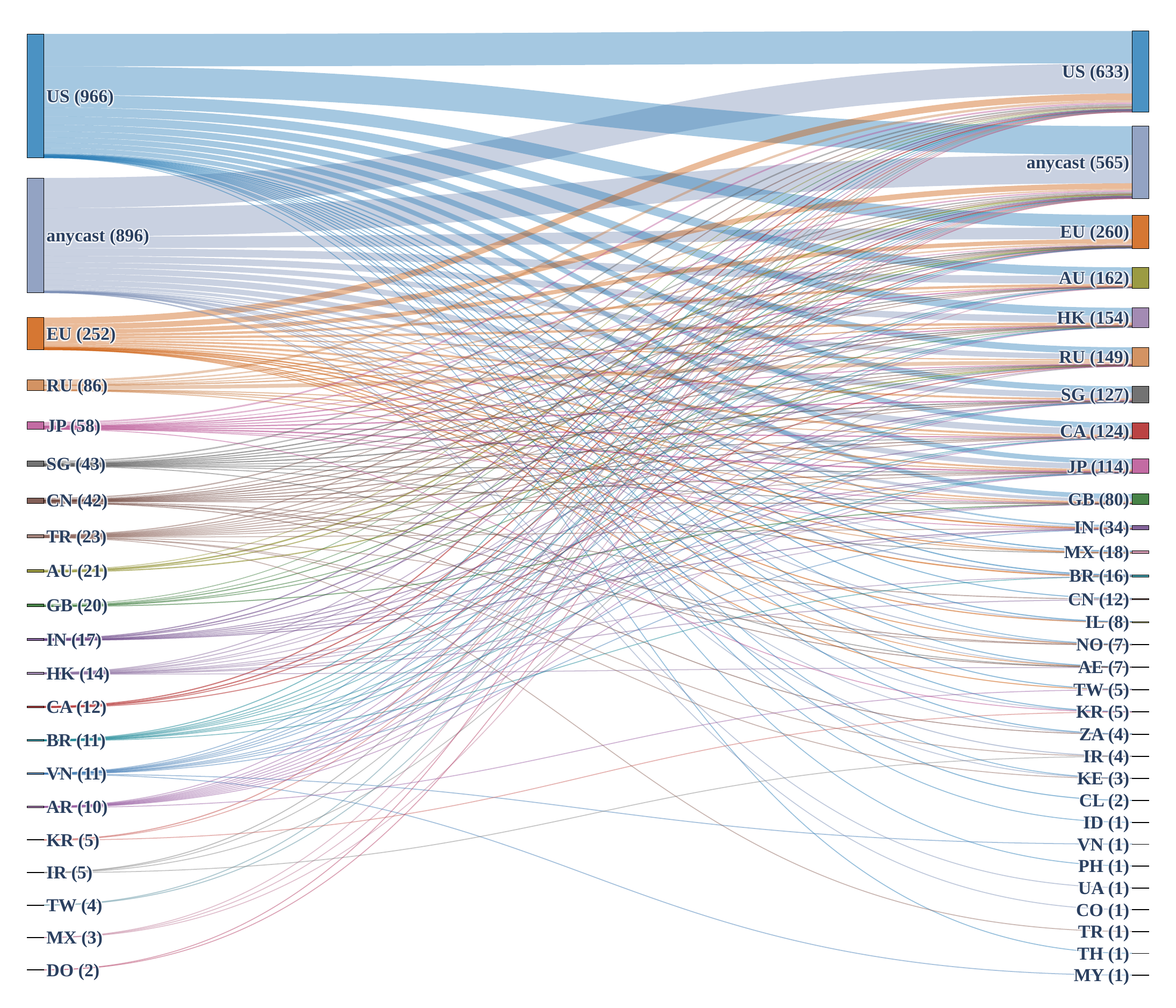}
		\vspace{-1cm}
		\caption{US-California}
		\label{fig:sankey_California}
	\end{subfigure}
	\begin{subfigure}[b]{0.48\textwidth}
		\centering
		\includegraphics[width=\textwidth]{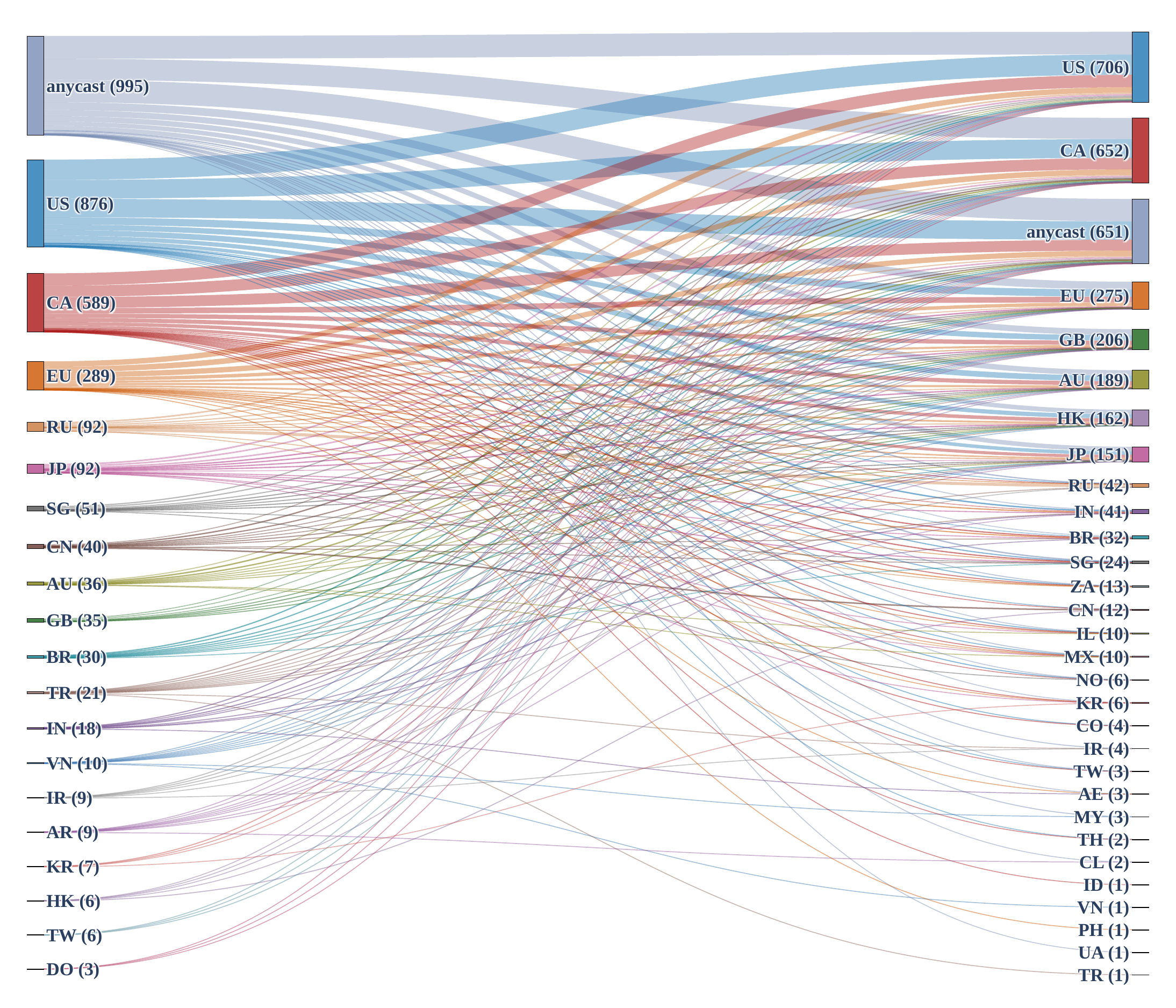}
		\vspace{-1cm}
		\caption{Canada}
		\label{fig:sankey_Canada}
	\end{subfigure}
	
	\begin{subfigure}[b]{0.48\textwidth}
		\centering
		\includegraphics[width=\textwidth]{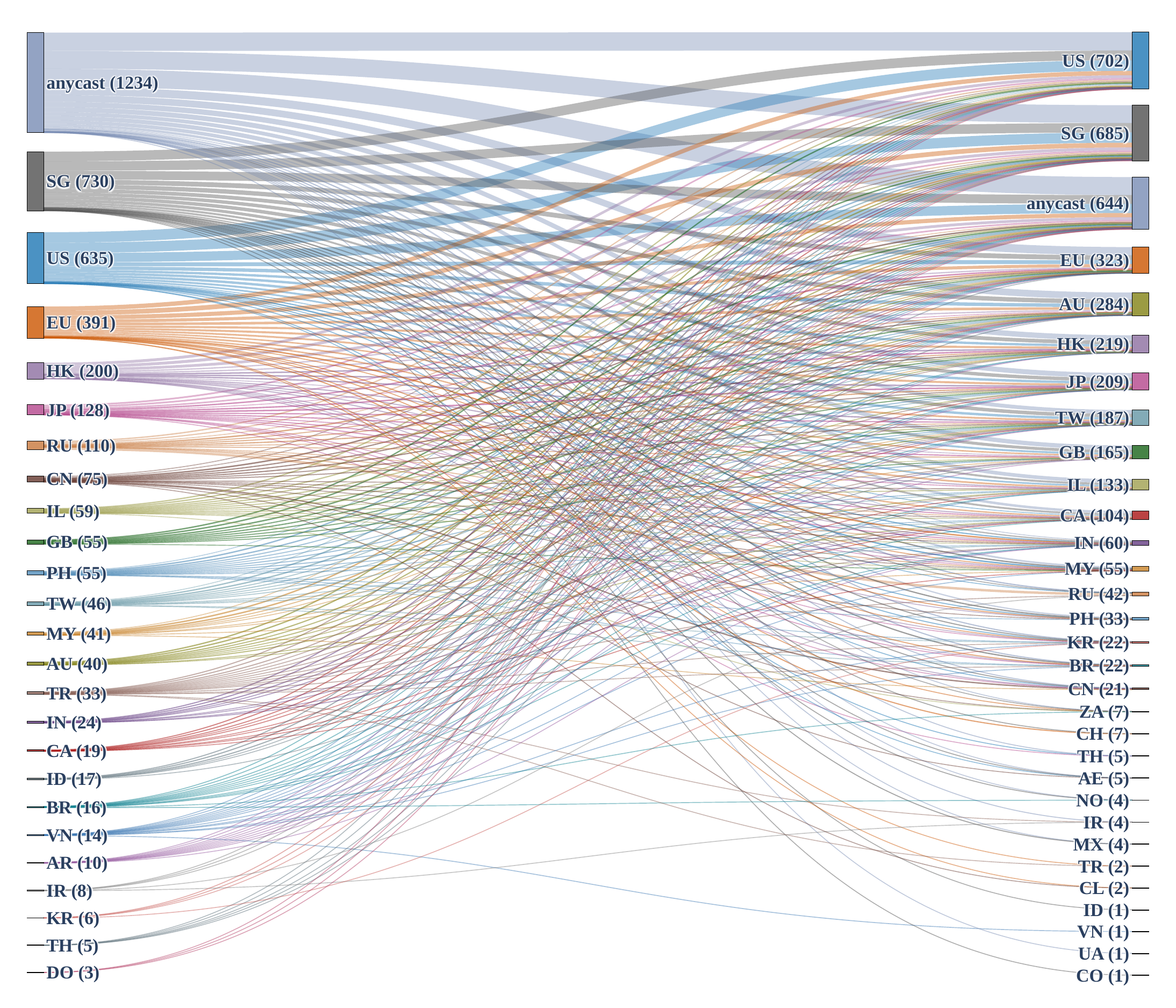}
		\vspace{-1cm}
		\caption{Singapore}
		\label{fig:sankey_Singapore}
	\end{subfigure}
	\begin{subfigure}[b]{0.48\textwidth}
		\centering
		\includegraphics[width=\textwidth]{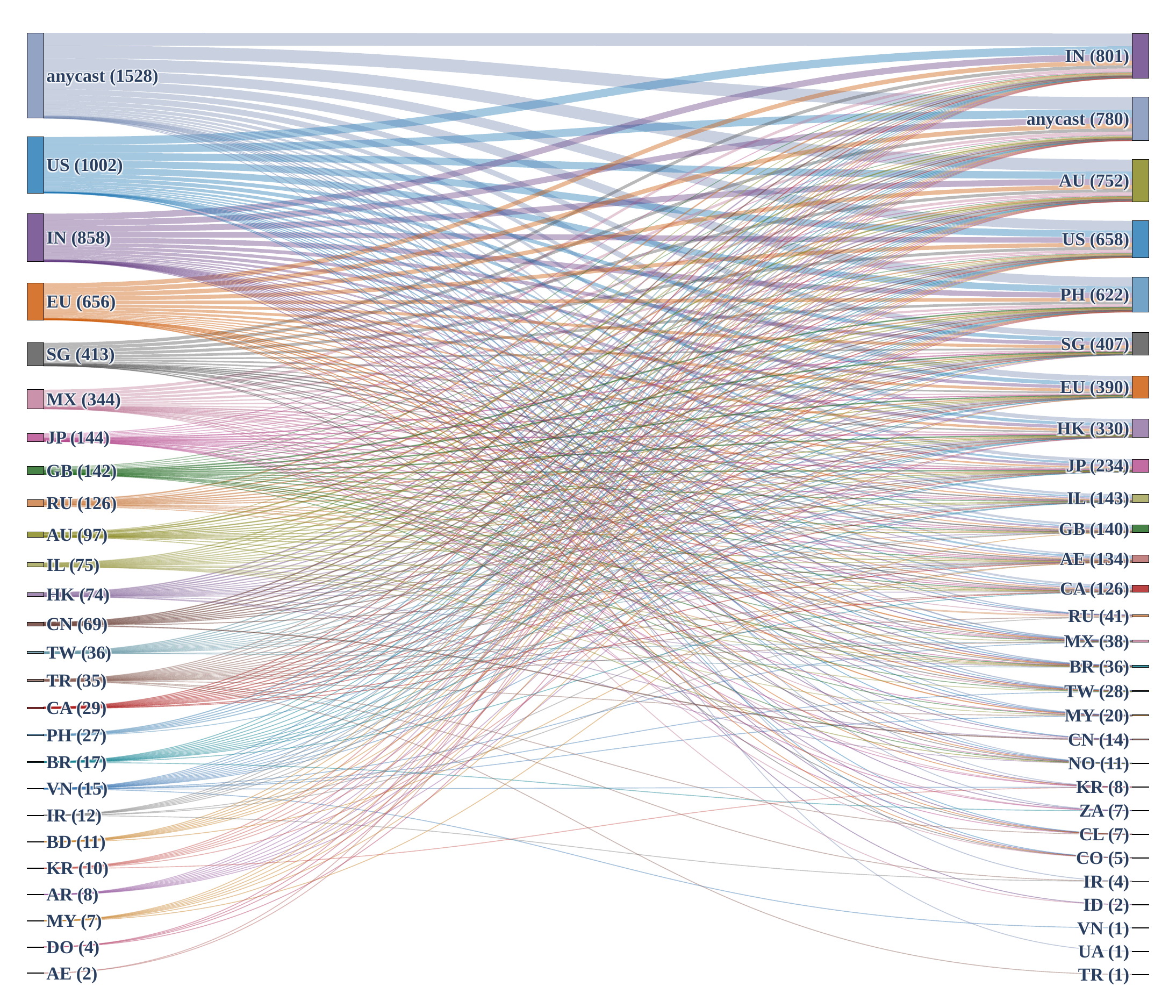}
		\vspace{-1cm}
		\caption{India}
		\label{fig:sankey_India}
	\end{subfigure}
	
	\caption{Cross-border data flows showing first-party (left) and \AT (right) server locations based on IP geo-location, across remaining six vantage points: Germany, Ireland, US-California, Canada, Singapore, and India.}
	\label{fig:sankey_all_vantages}
\end{figure*}

\autoref{fig:sankey_combined} in the main text shows cross-border data flows for Ohio and France. Here we present the corresponding diagrams for the remaining vantage points \autoref{fig:sankey_all_vantages}.

\clearpage
\subsection{Detailed Host Location Statistics}
\label{appendix:host-location}

\autoref{tab:host-location-cookie-banners} and \autoref{tab:host-location-cookie-banners-contribution} show respectively the prevalence of cookie banners and \AT contribution across host locations and vantage points.

\begin{table}[H]
	\centering
	\caption{Share of sites (in \%) that display a cookie banner, broken down by hosting jurisdiction (rows) and user vantage
		point (columns). EU vantages are grouped on the left, non-EU vantages on the right.}
	\label{tab:host-location-cookie-banners}
	\resizebox{\columnwidth}{!}{%
		\begin{tabular}{l|ccc|ccccc}
			\toprule
			& \multicolumn{3}{c|}{\textbf{EU vantages}} & \multicolumn{5}{c}{\textbf{Non-EU vantages}} \\
			\textbf{Host} & FR & DE & IE & OH & CA & CAD & IN & SG \\
			\midrule
			US(118) & \heatmap{61}\% & \heatmap{64}\% & \heatmap{63}\% & \heatmap{27}\% & \heatmap{31}\% & \heatmap{40}\% &
			\heatmap{33}\% & \heatmap{33}\% \\
			EU(55) & \heatmap{62}\% & \heatmap{60}\% & \heatmap{60}\% & \heatmap{42}\% & \heatmap{44}\% & \heatmap{44}\% &
			\heatmap{42}\% & \heatmap{42}\% \\
			SG(9) & \heatmap{33}\% & \heatmap{33}\% & \heatmap{33}\% & \heatmap{11}\% & \heatmap{11}\% & \heatmap{11}\% &
			\heatmap{11}\% & \heatmap{11}\% \\
			IN(4) & \heatmap{50}\% & \heatmap{50}\% & \heatmap{50}\% & \heatmap{25}\% & \heatmap{25}\% & \heatmap{25}\% &
			\heatmap{25}\% & \heatmap{25}\% \\
			CA(2) & \heatmap{50}\% & \heatmap{50}\% & \heatmap{50}\% & \heatmap{0}\% & \heatmap{0}\% & \heatmap{0}\% &
			\heatmap{0}\% & \heatmap{0}\% \\
			RU(33) & \heatmap{39}\% & \heatmap{39}\% & \heatmap{36}\% & \heatmap{30}\% & \heatmap{30}\% & \heatmap{24}\% &
			\heatmap{18}\% & \heatmap{24}\% \\
			CN(16) & \heatmap{19}\% & \heatmap{19}\% & \heatmap{19}\% & \heatmap{19}\% & \heatmap{19}\% & \heatmap{19}\% &
			\heatmap{19}\% & \heatmap{19}\% \\
			\bottomrule
		\end{tabular}%
	}
\end{table}

\begin{table}[H]
	\centering
	\caption{AT contribution \% from sites that present cookie banner in different regions}
	\label{tab:host-location-cookie-banners-contribution}
	\resizebox{\columnwidth}{!}{%
		\begin{tabular}{l|ccc|ccccc}
			\toprule
			& \multicolumn{3}{c|}{\textbf{EU vantages}} & \multicolumn{5}{c}{\textbf{Non-EU vantages}} \\
			\textbf{Host} & FR & DE & IE & OH & CA & CAD & IN & SG \\
			\midrule
			US(118) & \heatmap{81}\% & \heatmap{83}\% & \heatmap{82}\% & \heatmap{10}\% & \heatmap{17}\% & \heatmap{33}\% & \heatmap{18}\% & \heatmap{18}\% \\
			EU(55) & \heatmap{97}\% & \heatmap{97}\% & \heatmap{96}\% & \heatmap{55}\% & \heatmap{58}\% & \heatmap{59}\% & \heatmap{57}\% & \heatmap{56}\% \\
			SG(9) & \heatmap{90}\% & \heatmap{90}\% & \heatmap{88}\% & \heatmap{1}\% & \heatmap{1}\% & \heatmap{1}\% & \heatmap{2}\% & \heatmap{1}\% \\
			IN(4) & \heatmap{81}\% & \heatmap{82}\% & \heatmap{81}\% & \heatmap{6}\% & \heatmap{6}\% & \heatmap{5}\% & \heatmap{7}\% & \heatmap{9}\% \\
			CA(2) & \heatmap{80}\% & \heatmap{80}\% & \heatmap{82}\% & \heatmap{0}\% & \heatmap{0}\% & \heatmap{0}\% & \heatmap{0}\% & \heatmap{0}\% \\
			RU(33) & \heatmap{69}\% & \heatmap{69}\% & \heatmap{64}\% & \heatmap{35}\% & \heatmap{37}\% & \heatmap{33}\% & \heatmap{30}\% & \heatmap{34}\% \\
			CN(16) &  \heatmap{83}\% & \heatmap{92}\% & \heatmap{60}\% & \heatmap{76}\% & \heatmap{77}\% & \heatmap{89}\% & \heatmap{70}\% & \heatmap{75}\% \\
			\bottomrule
		\end{tabular}%
	}
\end{table}

\subsection{Concentration and Regional Clustering of Heavy Trackers}
We can also view this skewness from the perspective of \emph{who} contributes most of the \AT volume.
Given this heavy concentration, we next ask whether the same sites dominate everywhere or whether the identity of heavy trackers changes by region.
For each location, we take the top 10\% of sites by \AT contribution and measure the overlap between these sets across locations and show it in \autoref{tab:website-similarity1}.
We focus on the top 10\% because this subset captures the heaviest contributors while still leaving at least dozens of sites per location, which stabilizes overlap estimates; we observed qualitatively similar regional clustering when using other top-X\% thresholds.
We observe substantial overlap overall, suggesting a stable core of tracking-intensive sites that appear near the top of the ranking in many regions, but also clear regional clustering among the very heaviest contributors.
Our analysis shows that EU vantage points (FR, DE, IE) have a similarity of 0.9-0.92 among each other, while India and Singapore show 0.82 similarity with each other, which is higher than either EU or North America.
From North American vantage points, similarity is higher within North America than across regions; Asia also shows strong internal similarity.
This pattern is consistent with a picture in which a common global set of large sites dominates tracking, but their relative intensity and ranking vary by region, indicating location-aware advertising and analytics deployments.
\begin{table}[H]
	\centering
	\caption{Jaccard similarity of top 10\% websites contributing to third-party \AT across locations. High similarity within regional groups (NA, EU, Asia) indicates regional clustering of high-tracking websites.}
	\label{tab:website-similarity1}
	\resizebox{\columnwidth}{!}{%
	\begin{tabular}{l|ccc|cc|ccc}
		\toprule
		& \multicolumn{3}{c|}{\textbf{North America}} & \multicolumn{2}{c|}{\textbf{Asia}} & \multicolumn{3}{c}{\textbf{EU}} \\
		& CAD & OH & CA & IN & SG & FR & DE & IE \\
		\midrule
		CAD & \heatmappeachy{1.00} & \heatmappeachy{0.80} & \heatmappeachy{0.74} & \heatmappeachy{0.74} & \heatmappeachy{0.78} & \heatmappeachy{0.59} & \heatmappeachy{0.59} & \heatmappeachy{0.59} \\
		OH  & \heatmappeachy{0.80} & \heatmappeachy{1.00} & \heatmappeachy{0.80} & \heatmappeachy{0.74} & \heatmappeachy{0.76} & \heatmappeachy{0.57} & \heatmappeachy{0.57} & \heatmappeachy{0.56} \\
		CA  & \heatmappeachy{0.74} & \heatmappeachy{0.80} & \heatmappeachy{1.00} & \heatmappeachy{0.80} & \heatmappeachy{0.80} & \heatmappeachy{0.68} & \heatmappeachy{0.70} & \heatmappeachy{0.66} \\
		\midrule
		IN  & \heatmappeachy{0.74} & \heatmappeachy{0.74} & \heatmappeachy{0.80} & \heatmappeachy{1.00} & \heatmappeachy{0.87} & \heatmappeachy{0.70} & \heatmappeachy{0.66} & \heatmappeachy{0.68} \\
		SG  & \heatmappeachy{0.78} & \heatmappeachy{0.76} & \heatmappeachy{0.80} & \heatmappeachy{0.87} & \heatmappeachy{1.00} & \heatmappeachy{0.74} & \heatmappeachy{0.72} & \heatmappeachy{0.72} \\
		\midrule
		FR  & \heatmappeachy{0.59} & \heatmappeachy{0.57} & \heatmappeachy{0.68} & \heatmappeachy{0.70} & \heatmappeachy{0.74} & \heatmappeachy{1.00} & \heatmappeachy{0.90} & \heatmappeachy{0.92} \\
		DE  & \heatmappeachy{0.59} & \heatmappeachy{0.57} & \heatmappeachy{0.70} & \heatmappeachy{0.66} & \heatmappeachy{0.72} & \heatmappeachy{0.90} & \heatmappeachy{1.00} & \heatmappeachy{0.90} \\
		IE  & \heatmappeachy{0.59} & \heatmappeachy{0.56} & \heatmappeachy{0.66} & \heatmappeachy{0.68} & \heatmappeachy{0.72} & \heatmappeachy{0.92} & \heatmappeachy{0.90} & \heatmappeachy{1.00} \\
		\bottomrule
	\end{tabular}
	}
\end{table}

\subsection{ Chronological Summary of Related Work}
\label{app:rw}

Table~\ref{tab:rw-table} summarizes closely related studies along common study-design axes.
For each work, we report its primary objective, the number of browsers (or measurement client configurations), the \emph{browsing/measurement locations} from which measurements were conducted (i.e., vantage points), and the reported scale in number of sites (or the closest equivalent when the study uses a different primary unit).

\begin{table*}[p]
\centering
\caption{Regulatory and measurement-focused related work summarized along study-design axes. ``Geographic scope'' refers to \emph{browsing/measurement locations} (vantage points) used to collect data. ``\# Sites'' is reported as the number of sites (or the closest equivalent when the study uses observational user data).}
\label{tab:rw-table}
\small
\setlength{\tabcolsep}{4pt}
\begin{tabular}{p{1.8cm} p{0.7cm} p{5.5cm} p{1.2cm} p{4cm} p{1.1cm}}
\toprule
\textbf{Prior study} & \textbf{Year} & \textbf{Focus/Objective} & \textbf{Browsers} & \textbf{Geographic scope} & \textbf{\# Sites} \\
\midrule

Iordanou \hspace{1cm} et al.~\cite{iordanou2018tracing} &
2018 &
Cross-border tracking endpoints for EU users by mapping where third-party tracking communications terminate (destination infrastructure). &
1 (Chrome) &
Observational users (multi-country; EU28 subset analyzed; not controlled vantages). &
5,693 \\
\addlinespace

S{\o}rensen \hspace{1cm} et al.~\cite{sorensen2019before} &
2019 &
Third-party presence before vs.\ after GDPR enforcement using longitudinal crawling to quantify changes in third-party inclusion over time. &
1 (Firefox) &
Single EU-based browsing location (crawling VM in EU; not location-varied). &
1,250 \\
\addlinespace

Sanchez-Rola \hspace{1cm} et al.~\cite{sanchez2019can} &
2019 &
Tracking persistence after opt-out attempts: contrasts user-facing opt-out/consent choices with observed cookies and tracking activity. &
1 (Chrome) &
3 browsing locations (Spain; France; Ireland). &
2,000 \\
\addlinespace

Vallina \hspace{1cm} et al.~\cite{vallina2019tales} &
2019 &
Privacy practices and tracking in the adult-web ecosystem under GDPR, including measurement of tracking technologies and compliance signals. &
2 (Firefox and Chrome) &
Spain (physical) + VPN vantages in other EU member states + SG/IN/RU/US/UK. &
6,843 \\
\addlinespace

Urban \hspace{1.5cm} et al.~\cite{urban2020beyond} &
2020 &
Third-party dynamics ``in the field'': measures how third parties appear beyond landing pages and characterizes embedding patterns at scale. &
1 &
3 browsing locations (Europe/DE; North America/US; Asia/JP). &
10,000 \\
\addlinespace

Liu \hspace{1.5cm} et al.~\cite{liu2024opted} &
2024 &
Consent/CMP auditing under GDPR vs.\ CCPA contexts: tests whether opt-out/consent choices propagate to downstream advertising behavior. &
1 (Firefox) &
2 browsing locations (EU/Frankfurt; US/Northern California). &
352 \\
\addlinespace

Stafeev \hspace{1cm} et al.~\cite{stafeev2024sok} &
2024 &
Crawling methodology and measurement design space (SoK): evaluates how crawler strategy affects coverage and conclusions in web measurements. &
1 &
Not location-focused (no explicit browsing-location variation emphasized). &
2,000 \\
\addlinespace

Hantke \hspace{1cm} et al.~\cite{hantke2025web} &
2025 &
Web measurement accuracy and reproducibility: proposes recording/archiving and replay to support reproducible web archive construction and measurement fidelity. &
1 &
Not location-focused (tooling contribution; browsing location not a primary axis). &
10,000 \\
\addlinespace

Hausladen \hspace{1cm} et al.~\cite{hausladen2025websites} &
2025 &
CCPA/GPC opt-out compliance at scale and over time: evaluates whether sites honor Global Privacy Control signals in practice. &
1 (Firefox) &
Single browsing location (California via VPN; not location-varied). &
11,708 \\
\addlinespace

Singh \hspace{1.4cm} et al.~\cite{SinghIMC25} &
2025 &
Tracker exposure and related data flows in under-measured regions (Global South) using distributed, volunteer-based measurements across many countries. &
1 (Chrome) &
23 browsing locations (countries) across Africa/Asia/Europe/N.\ America/Oceania/S.\ America. &
$\approx$100 per country \\
\midrule

\textbf{\sysname} &
\textbf{2025} &
Cross-jurisdictional tracking under browser choice, user location, hosting jurisdiction, and consent state within a controlled factorial measurement design. &
\textbf{4} &
\textbf{8 browsing locations (NA/EU/Asia).} &
\textbf{743} \\
\bottomrule
\end{tabular}
\end{table*}

\end{document}